\documentclass[11pt,twocolumn]{article}

\usepackage{graphicx}
\usepackage{amsfonts}
\usepackage{amsmath}
\usepackage{amsbsy}
\usepackage{subfigure}
\usepackage{natbib}
\usepackage{color}

\def\JJ{{\bf {J}}}
\def\BB{{\bf {B}}}
\def\xx{{\bf {x}}}

\begin{document}

\title{Dynamics of braided coronal loops II: Cascade to multiple small-scale reconnection events}

\author{D.~I.~Pontin{$^*$}, A.~L.~Wilmot-Smith{$^*$}, G.~Hornig\footnote{Division of Mathematics, University of Dundee, Nethergate, Dundee, DD1 4HN, U.K.} ~\& K. Galsgaard\footnote{Niels Bohr Institute, Blegdamsvej 17, Dk-2100 Copenhagen {\O}, Denmark}}

\maketitle

\begin{abstract}
Aims: Our aim is to investigate the resistive relaxation of a magnetic loop that contains braided magnetic flux but no net current or helicity. The loop is subject to line-tied boundary conditions. We investigate the dynamical processes that occur during this relaxation, in particular the magnetic reconnection that occurs, and discuss the nature of the final equilibrium.
\\
Methods: The three-dimensional evolution of a braided magnetic field is followed in a series of resistive MHD simulations. \\
Results: It is found that, following an instability within the loop, a myriad of thin current layers forms, via a cascade-like process. This cascade becomes more developed and continues for a longer period of time for  higher magnetic Reynolds number. During the cascade, magnetic flux is reconnected multiple times, with the level of this `multiple reconnection' positively correlated with the magnetic Reynolds number. Eventually the system evolves into a state with no more small-scale current layers. This final state is found to approximate a non-linear force-free field consisting of two flux tubes of oppositely-signed twist embedded in a uniform background field.
\end{abstract}

\section{Introduction}
The braiding of magnetic loops in the solar corona, via convective motions at the solar surface, has long been suggested as a potential mechanism for heating the corona.  \cite{parker1972} proposed that in response to arbitrary footpoint motions at the photosphere, the coronal magnetic field will ideally relax towards a force-free equilibrium containing tangential discontinuities, corresponding to current sheets. Therefore, it was argued, as such a singular state is approached during the relaxation, the diffusivity of the plasma must always become important, leading to magnetic reconnection and plasma heating. Since this idea was first proposed, there have been many arguments for and against its validity \citep[e.g.][]{vanballegooijen1985,longcope1994,ng1998}.

In Parker's original `topological dissipation' model, an initially uniform magnetic field is taken between two perfectly conducting parallel plates. Then random (smooth) motions are applied on the perfectly conducting plates. A number of attempts have been made to simulate such a scenario numerically, using a variety of numerical approaches. \cite{craigsneyd2005}  employed an ideal relaxation scheme which includes a fictitious frictional term in the equation of motion \citep{craig1986}. They applied various complex deformations at the boundaries of the domain. They found that regardless of the nature or extent of the deformation, no tangential discontinuities developed in the relaxed state, with the current concentrations remaining large-scale in all cases. \cite{mikic1989} employed a slightly different approach. They began with a uniform field and then sequentially applied large-scale shear flows of random orientation on one boundary (while the field at the other boundary was held fixed). In the time between each boundary shear, they solved the ideal MHD equations { (neglecting the pressure and advective terms in the equation of motion)}, including a large spatially-uniform viscosity in order to relax the magnetic field towards equilibrium. They found, once again, that no discontinuities were formed during any of these relaxation processes. However, progressively smaller scale current structures were found to develop as the number of shear disturbances increased. The implication is therefore that after a sufficient length of time, the scales would reach those appropriate for dissipation in the corona. They found, furthermore, that the current density in the domain increased exponentially in time, which is consistent with the earlier analysis of \cite{vanballegooijen1985}, who predicted thin non-singular current layers rather than tangential discontinuities.

{\cite{galsgaard1996} employed a further different approach to those discussed above, solving the full set of resistive MHD equations without employing any artificial force (such as enhanced viscosity) to inhibit the plasma dynamics}. They applied a similar sequence of large-scale shearing motions as \cite{mikic1989}, this time at both driving boundaries. The amplitude and orientation of the shearing were chosen at random (from a normal distribution). The results were similar to those of \cite{mikic1989}, in that after only a few Alfv{\' e}n crossing times, multiple small-scale current filaments were found to form in the domain. However, with the shearing applied on both boundaries, exponentially growing currents were already obtained {after the second shear motion due to the interlocking of the field lines, with the tension force setting up a stagnation flow}. Due to the finite resistivity in their simulations, magnetic reconnection and Joule dissipation occurred in these filaments. This energy release was found to be intermittent or `bursty' when the time scale for energy input (via the boundary shearing) was long compared with the Alfv{\' e}n crossing time. The work was extended by  \cite{galsgaard2002} with the inclusion of a stratified atmosphere.
{Furthermore, the generation and properties of turbulence in Parker's model have been studied in the framework of reduced MHD by a number of authors. In the initial studies, two-dimensional simulations were performed, with imposed `forcing terms' taking the place of the boundary driving \citep[e.g.][]{einaudi1996,dmitruk1997,georgoulis1998}. More recently, three-dimensional simulations of Parker's model with the reduced MHD equations have been performed \citep[e.g.][]{dmitruk2003,rappazzo2008}, and the resulting energy spectra and heating event statistics investigated.}

In the present work, we approach the topological dissipation problem discussed above from a different angle to previous studies. 
{We take a braided magnetic field that is close to force-free as an initial condition,
and concentrate on the details of the subsequent evolution, including current sheet formation.
In general, braided force-free fields could arise in the solar corona either through the emergence of braided
flux from the interior or through random footpoint motions at the solar surface, accompanied by some 
intermediate relaxation processes.
The ideal relaxation of this field towards a force-free state was considered in \citet{wilmotsmith2009a}
where only large-scale current features were found in the end state.
}
We now follow the evolution of the system in a resistive MHD simulation. In a previous paper (\cite{wilmotsmith2010}, hereafter referred to as Paper I) we described how during the early evolution an instability occurs which moves the system away from equilibrium, leading to the formation of current sheets and thus the onset of magnetic reconnection. In the present paper we address the following key questions:
\begin{enumerate}
\item
As the plasma seeks a new equilibrium, what is the nature of the resistive relaxation process? 
\item
In particular what are the properties of the magnetic reconnection processes that take place?
\item
What is the nature of the final state of the relaxation?
\item
What is the dependence of each of the above on the magnetic Reynolds number?
\end{enumerate}

In Section \ref{numsec} we introduce the numerical method and summarise the results of our previous investigations. In Section \ref{evsec} we describe qualitatively the evolution of the system, and in Section \ref{finsec} we discuss the nature of the final state. In Section \ref{recsec} we investigate the reconnection in the system  in more detail, and then in Section \ref{etasec} discuss the dependence on the plasma resistivity. Finally in Section \ref{concsec} we present our conclusions.

\section{Numerical method and early evolution}\label{numsec}
\subsection{Numerical scheme}
The numerical scheme employed in the simulations that follow is described
briefly below (further details may be found in \cite{nordlund1997} and at
http://www.astro.ku.dk/$\sim$kg). We solve the three-dimensional resistive MHD
equations in the form
\begin{eqnarray}
\frac{\partial {\bf B}}{\partial t} & = & - \nabla \times {\bf E},
\label{numeq1}\\ 
{\bf E} & = & -\left( {\bf v} \times {\bf B} \right)
\: + \: \eta {\bf J}, \label{numeq2}\\ 
{\bf J} & = & \nabla \times
{\bf B}, \label{numeq3}\\ 
\frac{\partial \rho}{\partial t} & = & -
\nabla \cdot \left( \rho {\bf v} \right), \label{numeq4}\\
\frac{\partial}{\partial t}\left( \rho {\bf v} \right) & = & - \nabla
\cdot \left( \rho {\bf v} {\bf v} \: + \: {\underline {\underline
\tau}} \right) \: - \: \nabla P \: + \: {\bf J} \times {\bf B},
\label{numeq5}\\ 
\frac{\partial e}{\partial t} & = & -\nabla \cdot
\left( e {\bf v} \right) \: - \: P \: \nabla \cdot {\bf v} \: + \:
Q_{visc} \: + \: Q_{J} \label{numeq6},
\end{eqnarray}
where ${\bf B}$ is the magnetic field, ${\bf E}$ the electric field,
${\bf v}$ the plasma velocity, ${\rho}$ the density, $\eta$ the resistivity, ${\bf J}$ the
electric current density, ${\underline {\underline
\tau}}$ the viscous stress tensor, $P$ the pressure, $e$ the thermal
energy, $Q_{visc}$ the viscous dissipation and $Q_{J}$ the Joule
dissipation. An ideal gas is assumed, and hence $P \: = \: \left(
\gamma -1 \right) \: e \: = \: {\textstyle \frac{2}{3}}e$.
The equations above have been non-dimensionalised by setting
the magnetic permeability $\mu_0 = 1$, and the gas constant equal to
the mean molecular weight. The result is that for a volume in which $| \rho |=| {\bf B} | = 1$, time units are such that an Alfv\'{e}n wave would travel one space unit in one unit of time. 

In all simulation runs we employ an explicitly prescribed, spatially uniform resistivity, and thus we have simply $Q_J=\eta J^2$. Viscosity is calculated using a combined second-order and fourth-order method (sometimes termed `hyper-viscosity'), which is capable of providing sufficient localised
 dissipation where necessary to handle the development of numerical
 instabilities \citep{nordlund1997}. The effect is to `switch on' the viscosity where very short length scales develop in ${\bf v}$, while maintaining a minimal amount of viscous dissipation where the velocity field is smooth. {As such, the total viscous dissipation in the domain is negligible when compared with the Joule dissipation.} 
All simulations are carried out on a grid with $512^3$ nodes. The dimensionless plasma density  is initialised to have value $\rho=1$ throughout the domain, while the thermal energy is initialised with value $e=0.1$. The result is a plasma-$\beta$ that varies throughout the domain between 0.10 and 0.14 at $t=0$. 
 
{We have checked our numerical results by repeating the standard run described below (with $\eta=10^{-3}$) using the {\tt Lare3d} MHD code \citep{arber2001}. The qualitative results described in sections \ref{evsec}--\ref{etasec} have been verified with this different scheme, demonstrating their robustness.}

\subsection{Initial conditions and early time evolution}
\begin{figure}[]
\centering
(a)\includegraphics[height=9cm]{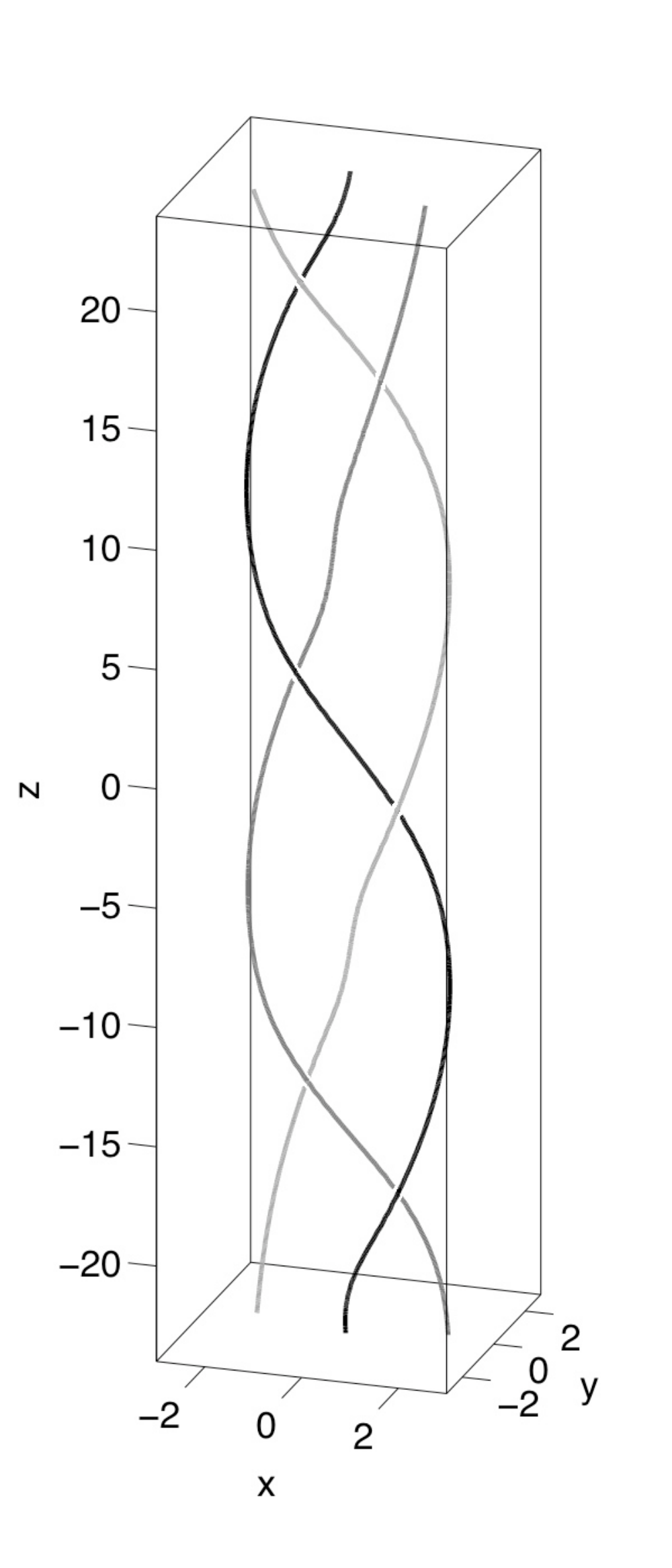}~~
(b)\includegraphics[height=9cm]{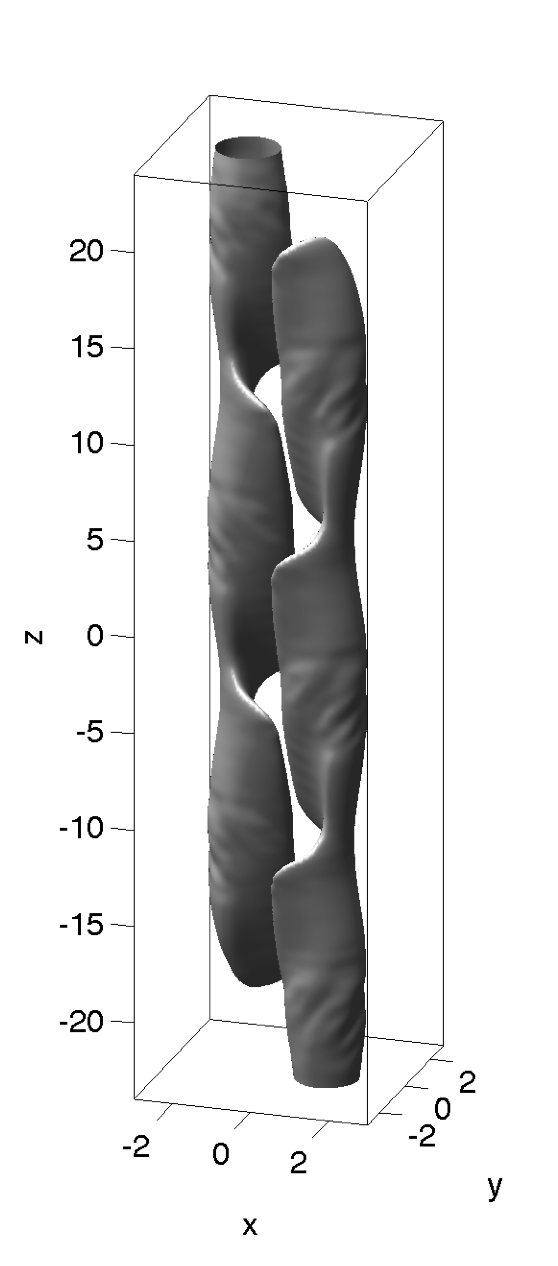} 
\caption{(a) Three representative magnetic field lines from the initial magnetic field ($t=0$) which demonstrate the braiding present in the field. (b) Isosurface of current density $|\JJ|$ at $t=0$, at 25\% of its maximum value.}
\label{initfig}
\end{figure}

We use as an initial condition for our resistive MHD simulations a magnetic field that is close to being force-free  \citep[][]{wilmotsmith2009a}, and that is obtained via an ideal Lagrangian relaxation scheme \citep{craig1986, pontin2009}. This is explained in detail in Paper I. In what follows, we term the input field for the ideal relaxation scheme the `pre-initial' magnetic field. The final state of this relaxation, which is the initial condition for the resistive MHD simulations described herein, is termed the `initial' magnetic field. The pre-initial and initial magnetic fields have the same braided magnetic topology (as the relaxation is exactly ideal). The magnetic field is modelled on the so-called pigtail braid, with a subset of the field lines being linked in a non-trivial way -- see Figure \ref{initfig}(a). The pre-initial structure consists of six localised regions of twisted flux in an otherwise homogeneous field. These twists are of equal magnitude, but with opposite sign in three of the regions, such that the net current density and net helicity in the computational volume are both zero. (To be precise we calculate the relative helicity \citep{bergerfield1984} using the potential field (${\bf B}_p=[0,0,1]$) as the reference field. In our case we can choose a gauge such that ${\bf A}_p = {\bf A}$ on the boundary so that the relative helicity reduces to an integral over ${\bf A}\cdot{\bf B}$ only.) In the initial magnetic field used for our MHD simulations, there are two large-scale, diffuse current structures within the domain, but no thin current layers -- see Figure \ref{initfig}(b) and \cite{wilmotsmith2009a,wilmotsmith2009b} for a full description. 

The numerical domain has dimensions $[x,y,z]\in [\pm6, \pm6, \pm24]$.  The magnetic field is line-tied on $z=\pm 24$,  where the plasma velocity is fixed at zero for the whole experiment, while the $x$ and $y$ boundary conditions are periodic. {Since $\BB=[0,0,1]$ at the $x$- and $y$-boundaries, these boundaries are closed with respect to the magnetic field at $t=0$}. We find that the dynamics is confined within the domain ($x,y \in [-4,4]$) and, with the confining background field, field lines do not leave the numerical box through the side boundaries at any stage of the simulation. 
We note finally that  the simulations are run for a sufficient time that in the final state the resistive dissipation is occurring on the global length scale of the regions of twist that were present in the pre-initial field (see below).

The evolution of the system at early stages can be summarised as follows, and is described in detail in Paper I. The system initially remains relatively stationary, being close to a force-free equilibrium. However, at approximately $t=8$ an instability sets in, and the current intensifies in two thin layers. {At present the nature of this instability is not fully understood, though we propose that it may be a resistive instability analogous to the internal kink instability.} This time interval corresponds to the Alfv{\' e}n travel time between two of the imposed regions of twist in the pre-initial magnetic field. We go on below to investigate the continued evolution of the system.

\section{Long time evolution}\label{evsec}
\subsection{Qualitative behaviour}\label{waffle}
\begin{figure*}
\centering
\includegraphics[width=0.193\textwidth]{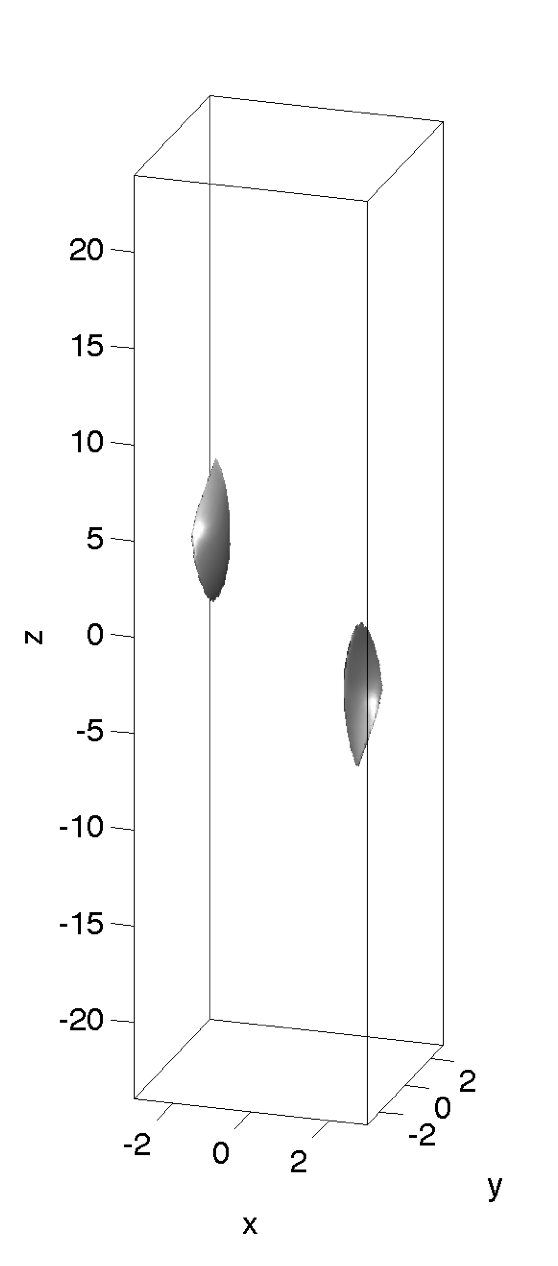}
\includegraphics[width=0.193\textwidth]{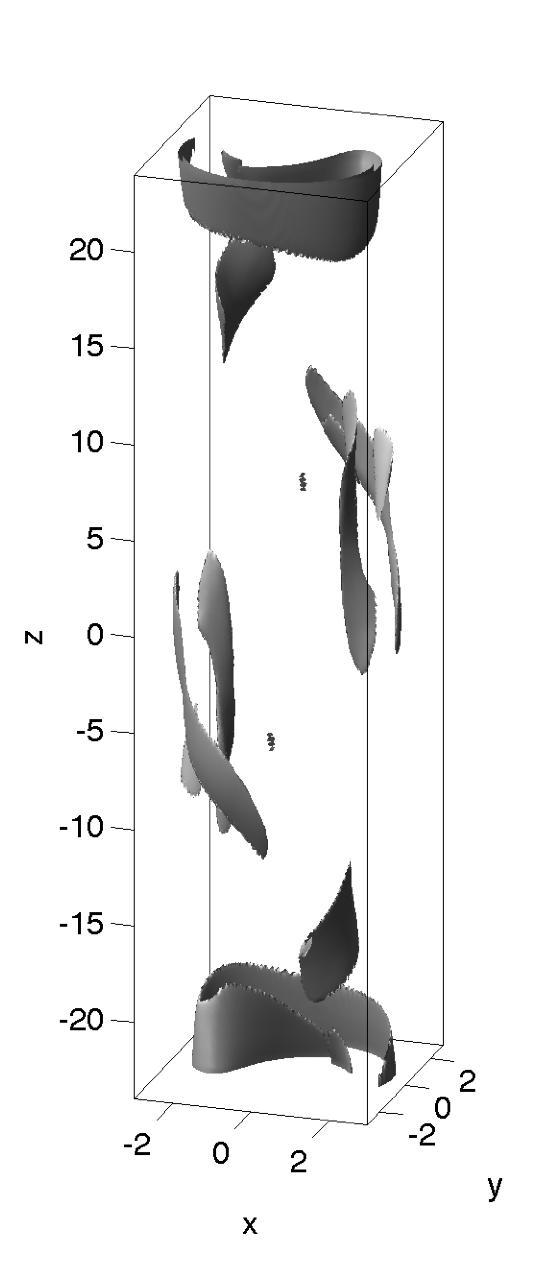}
\includegraphics[width=0.193\textwidth]{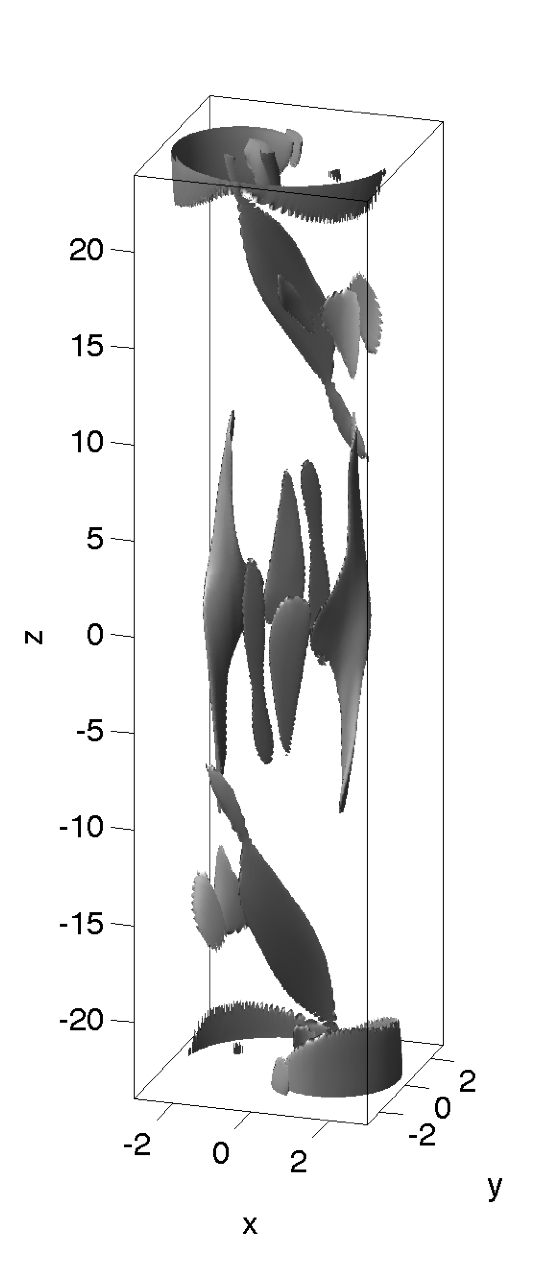}
\includegraphics[width=0.193\textwidth]{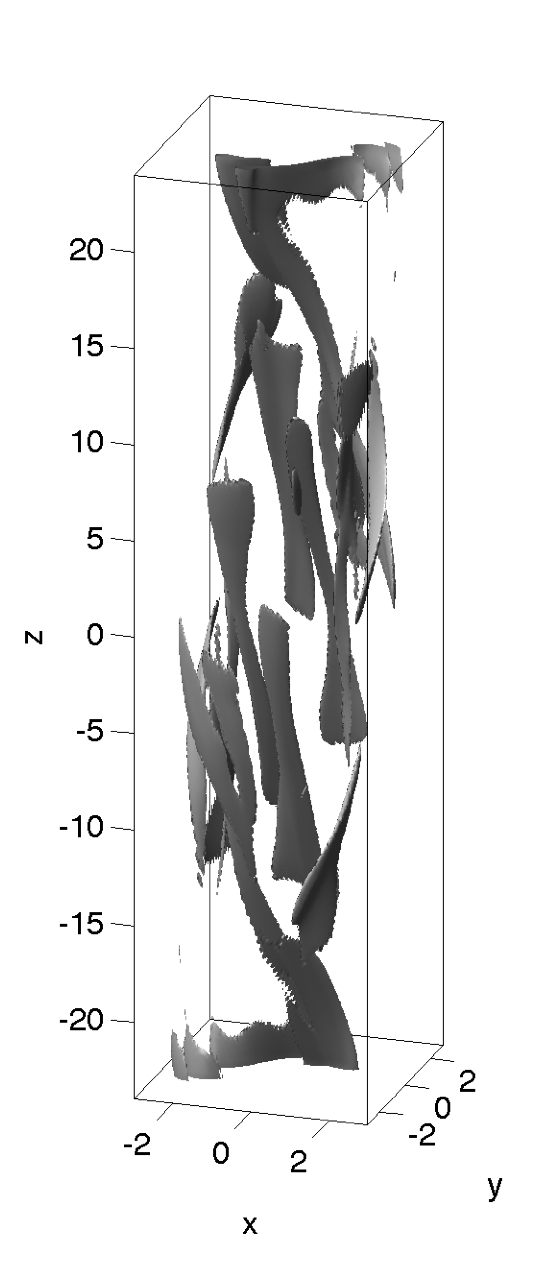}
\includegraphics[width=0.193\textwidth]{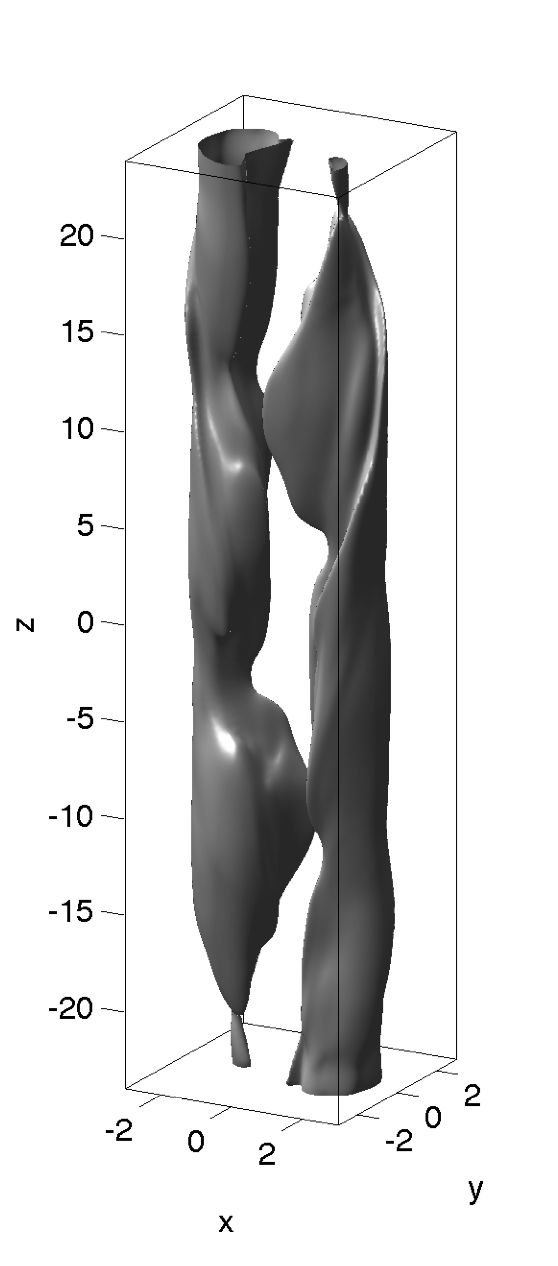}
\caption{Isosurfaces of the current density $|\JJ|$ at $t=15, 27.5, 35, 50, 290$, at 50\% of the corresponding maximum value of $|\JJ|$. For the run with $\eta=10^{-3}$.}
\label{jiso}
\end{figure*}

{Following the instability which gives rise to the initial current layers, the system goes on to develop an increasing number of thin current layers,  
as illustrated in Figure~\ref{jiso} where isosurfaces of current density at 50\% of the domain 
maximum are shown.
It is worth noting that the two initial current layers  are stronger than any of those that form in the subsequent evolution (see the following section and Figure \ref{relaxfig}). The development of a highly fragmented current structure means that magnetic reconnection also occurs in a highly fragmented portion of the domain. This magnetic reconnection is investigated in detail in Section \ref{recsec}. The evolution resembles qualitatively the onset of a turbulent-like behaviour, with the current layers forming a highly disordered pattern which fills the central portion (in $xy$) of the computational domain (the region in which the initial field line mapping is non-trivial). 
Investigating the power spectra of relevant physical quantities to ascertain whether indeed true turbulence develops is beyond the scope of this paper. It is in any case likely to require higher numerical resolution (and the lower value of imposed $\eta$ that this permits) in order to resolve a sufficient number of decades of spatial scales. Thus, in the remainder of this paper we use the terms turbulence and cascade only in a very loose sense.}
\begin{figure}
\centering
\includegraphics[width=0.48\textwidth]{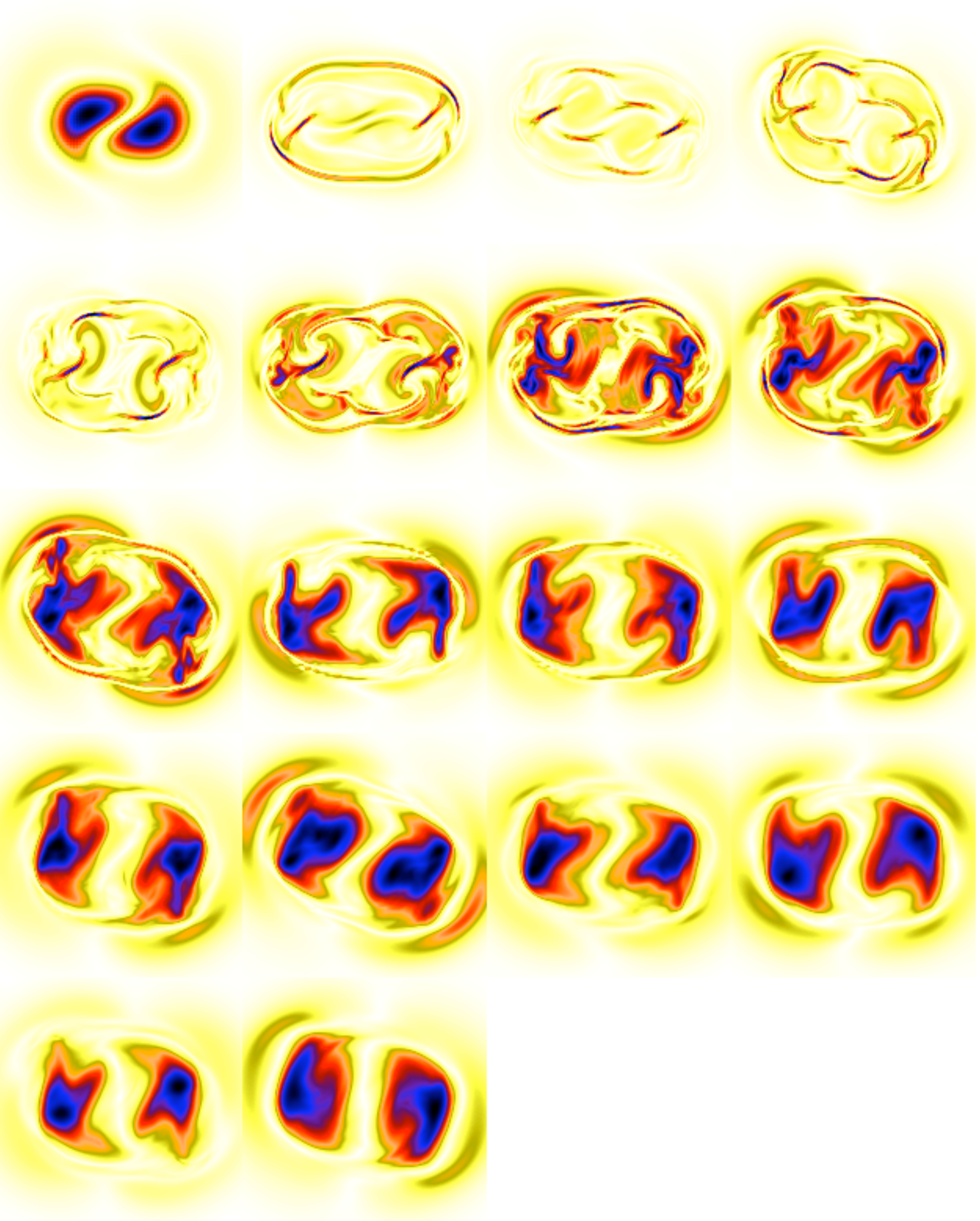}
\caption{Current density $|\JJ|$ at $z=0$ for $[x,y]\in [\pm 3.5, \pm 3.5]$ every $t=20$ units from $t=0$ to $t=340$ (top-left to bottom-right). The shading intensity is scaled individually to the maximum at each time, for clarity. For the run with $\eta=10^{-3}$.}
\label{jz0}
\end{figure}

In Paper I, it was shown that the locations where the first pair of current sheets form are well predicted by analysing in combination the squashing factor $Q$ of the field line mapping \citep{titov2002,titov2007} and the integrated parallel current along field lines, $\mathcal{J}=\int J_\| dl$. The initial current layers form in regions threaded by field lines with high values of both quantities (for an explanation of the importance of $\mathcal{J}$ see \cite{wilmotsmith2009a}). The question naturally arises: do the secondary, tertiary and subsequent current layers form in response to the dynamics that are set in motion by the initial instability and reconnection process? 
Or are they the result of multiple instabilities which occur throughout  the  domain with different growth rates -- with the initial current sheets that we observe being simply those associated with the fastest growing instabilities? The myriad of thin layers of high $Q$ and $\mathcal{J}$ found in the domain would support the latter idea. 
The evidence suggests that neither of these alternatives is solely responsible, but rather a combination of the two. However, this is another aspect of the simulations that warrants further investigation.

In an attempt to gain some insight, one can look at individual reconnection events in the domain. {One can identify the outflow from certain current layers with the inflow to another reconnection process (this being one 
typical picture of how  {a multitude of current layers} may form,
where one reconnection process drives the next and so on \cite[e.g.][]{galsgaard1996,watson2003,hood2009}). However, there are many other current layers for which this is clearly not the case. Moreover,} determining the local relationship between reconnection processes is not straightforward -- in three dimensions reconnection need not necessarily occur in a region of hyperbolic field, and is also not always associated with a hyperbolic flow structure with well-defined inflow and outflow. In fact, in three dimensions the characteristic property  for reconnection in the absence of null points is a counter-rotational plasma flow on either side (with respect to the direction of ${\bf B}$) of the diffusion region \citep{hornig2003}. Thus, determining the dynamic interaction between different reconnection processes within the volume is a non-trivial problem.

The  increase in the complexity of the current distribution  continues until around $t=80$, after which the number of current layers begins to decrease. Eventually, the current concentrations again attain large scales (see Figures \ref{jiso} and \ref{jz0}). The properties of the final state of the simulation are considered in detail in Section \ref{finsec}. The extent and duration of the fragmentation of the current layers -- and the subsequent time taken to return to a configuration with currents on system-size scales -- is found to be dependent on the value taken for the resisitvity, $\eta$. This is discussed in Section \ref{etasec}.

\subsection{Relaxation properties}
The magnetic field begins in a state that is close to a (non-linear) force-free equilibrium. As mentioned above (and described in more detail in Paper I), an instability is then triggered forcing the system away from equilibrium. The subsequent resistive evolution can be described as a relaxation process, which is mediated by magnetic reconnection (see Section \ref{recsec}). 
\begin{figure}
\centering
\includegraphics[width=0.5\textwidth]{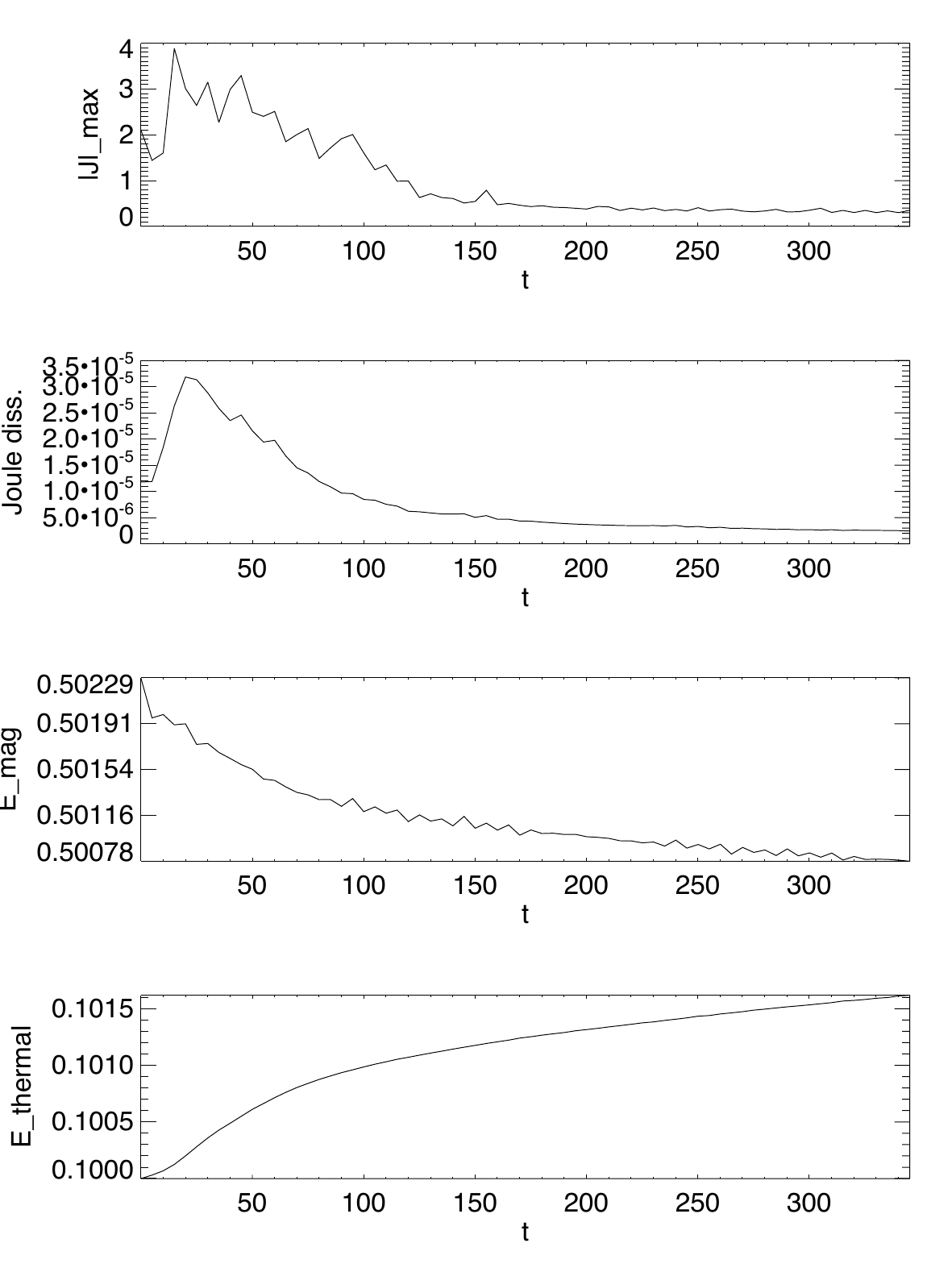}
\caption{From top to bottom: Time evolution of the spatial maximum of $|\JJ|$; the Joule dissipation $\eta J^2$ per unit volume, the magnetic energy per unit volume, and the thermal energy per unit volume, for the run with $\eta=10^{-3}$.}
\label{relaxfig}
\end{figure}
\begin{figure}
\centering
\includegraphics[width=0.45\textwidth]{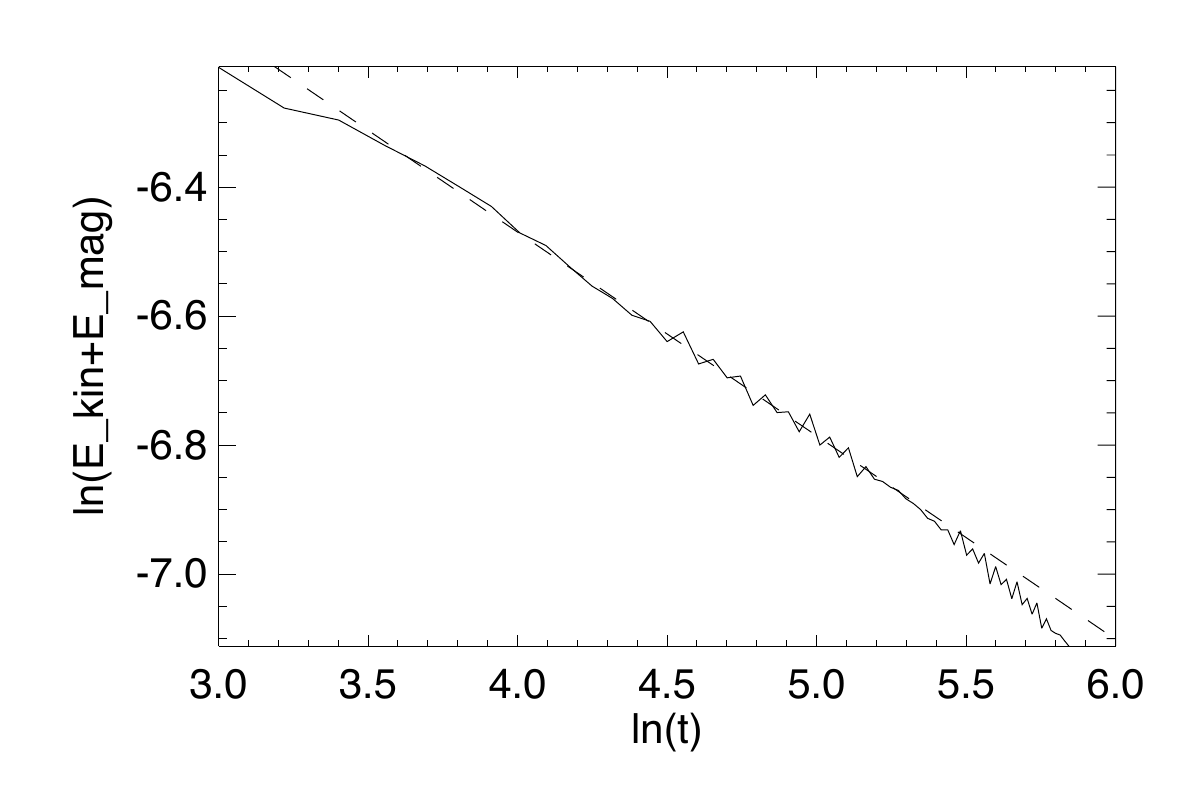}
\caption{Total per unit volume of the kinetic energy plus the magnetic energy in excess of that of the background field,  plotted on a $\log$-$\log$ scale, for the run with $\eta=10^{-3}$.}
\label{etotloglog}
\end{figure}
{This relaxation process is demonstrated in Figure \ref{relaxfig}
{where the evolution of various quantities with time is shown.}
The top panel shows  a time series of the peak current in the domain. After an initial sharp drop (that is likely due largely to the change from one discretisation to another), this soon peaks (at $t=12$, see Paper I) before gradually declining towards the end of the simulation run. 
{The current settles to a constant value after $t \approx 200$.  After this point the configuration is approximately stationary, although the simulation has been run to $t=350$ to be sure we are in a state whose topology is stable.}
The Joule dissipation per unit volume  shows a similar pattern of evolution to the peak current. 
{The free magnetic energy in the initial state is relatively modest, with only 0.5\% magnetic energy per unit volume above that of the potential (homogenous) field.} As time progresses, this magnetic energy decays, being converted into kinetic and  thermal energy (see the lower panel of the figure). We note that in the final state, the magnetic energy  is still around $0.2\%$ above that of the homogeneous field. In total over 60\% of the free energy in the initial configuration is converted to other forms. The nature of the equilibrium {approached} 
is discussed in detail in the following section.}

As noted above the relaxation process has a turbulent appearance. However, due to the moderate Reynolds numbers that we are able to use with the present resolution, the system is unlikely to be in a state of fully developed turbulence, though we anticipate that such a state would be reached were we able to perform simulations with lower values of the resistivity. In terms of the categorisation described by \cite{biskamp2003}, the system would then be described as undergoing decaying turbulence, since the additional energy in the system is included as excess magnetic energy in the system at large scales at $t=0$ (in contrast with the forced turbulence simulations in which a mechanical stressing is continually applied, e.g.~\cite{einaudi1996,dmitruk1997,georgoulis1998,dmitruk2003,rappazzo2008}). The sum of the magnetic and kinetic energies is expected to decay exponentially in time during decaying turbulence \citep{biskamp2003}. In Figure \ref{etotloglog} this quantity is plotted versus time on a log-log scale, where we have subtracted the constant contribution to the magnetic energy arising from the uniform background field, for clarity. The dashed line shows a least-squares fit for the data points between $t=20$ and $t=180$ when the current distribution contains small scales. It is clear that we do indeed see a power law dependence during this time, which is suggestive that we are at least approaching a regime of decaying turbulence. It is also apparent from the plot that after $t\approx 180$ (after which physical quantities vary on system-size scales), this power law is no longer followed, with a steepening of the curve suggesting a possible transition towards an exponential decay. 

While the simulation has been run to $t=350$, it is found that after $t \approx 200$ the topology of the magnetic field changes only due to diffusion in the two tubular current structures which have system-size dimensions both along ${\bf B}$ and perpendicular to ${\bf B}$ (see Figures \ref{jiso} and \ref{jz0}, and Figure \ref{phineg_sd} bottom panel). Owing to the enormous diffusive timescale in the corona, this diffusive evolution is not relevant for the coronal field. We therefore consider the state reached at the end of our simulations to characterise the physically relevant final state.  We therefore take $t=290$ as the final state of the simulation for our investigations into its nature that follow. It should be noted that this final state is not an exact force-free equilibrium, due to the finite plasma pressure, as well the residual Alfv{\' e}nic oscillation of the structure about its equilibrium (our use of a hyper viscosity means that these oscillations are damped only very slowly). However, it is close to force-free, in that in terms of the dimensionless measure of the proximity to a force-free field, $\epsilon l$, given in \citet{wilmotsmith2009a}, we have $\epsilon l \approx 6 \times 10^{-3}$ (where $\epsilon l \in [0,1]$ and $\epsilon l =0$ for a perfectly force-free field).

 \section{Nature of the final state}\label{finsec}

\subsection{Properties of the relaxation}

\begin{figure}[]
\begin{center}
\includegraphics[width=0.42\textwidth]{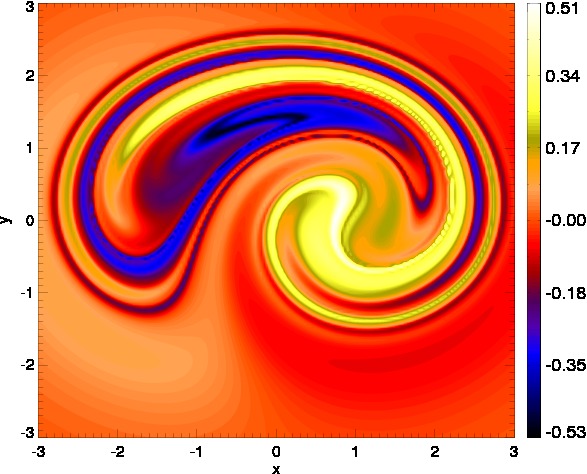}
\includegraphics[width=0.42\textwidth]{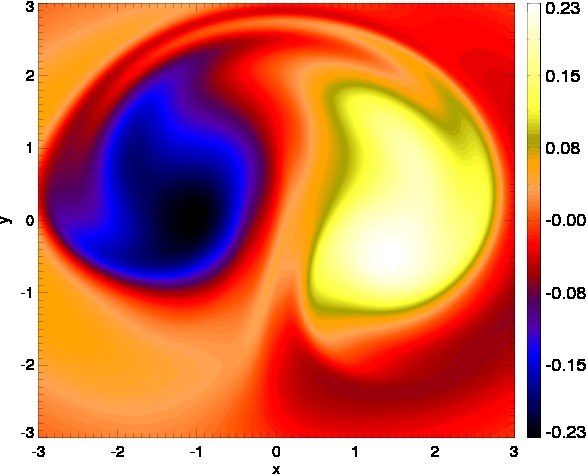}
\end{center}
\caption{The mean value of $\alpha^{*}$ along field lines shown on a section of the lower
boundary ($z=-24$) of the domain in the initial ({\it top}, $t=0$) and final ({\it bottom}, $t=290$)
states showing a smoothing of $\alpha^{*}$ during the resistive relaxation with $\eta=10^{-3}$.  
}
\label{fig:alpha}
\end{figure}

In order to investigate the nature of the final state, we take  a grid of starting points on the lower boundary $z=-24$ (taking $160^{2}$ points over the region $[-3,3]^{2}$) 
and trace field lines through the volume to the upper boundary.   For each field line, we calculate  ${\overline{ \alpha^{*}}}$, the mean value along the field line of the quantity
$ \alpha^{*} = (\mathbf{J} \cdot \mathbf{B})/(\mathbf{B}\cdot\mathbf{B})$.
Note that for a perfectly force-free field this simply reduces to the force-free parameter $\alpha$
(where $\mathbf{J} = \alpha \mathbf{B}$) which is constant along field lines.
Figure~\ref{fig:alpha} shows contours of  $\overline{ \alpha^{*}}$ found in this way for the initial state
(top panel) and the final state ($t=290$, lower panel).  In both cases the sum of 
${\overline{ \alpha^{*}}}$  over all field lines in the domain is zero.
In the final state the small scales in ${\overline{ \alpha^{*}}}$ have been smoothed out leaving two 
large-scale patches of opposite signs.   Note in addition that the extremes of the value of ${\overline{ \alpha^{*}}}$ 
have been reduced.

\begin{figure}[]
\begin{center}
\includegraphics[width=0.24\textwidth]{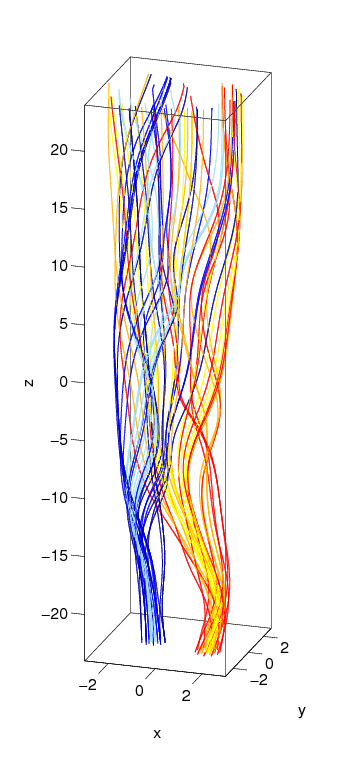}
\includegraphics[width=0.24\textwidth]{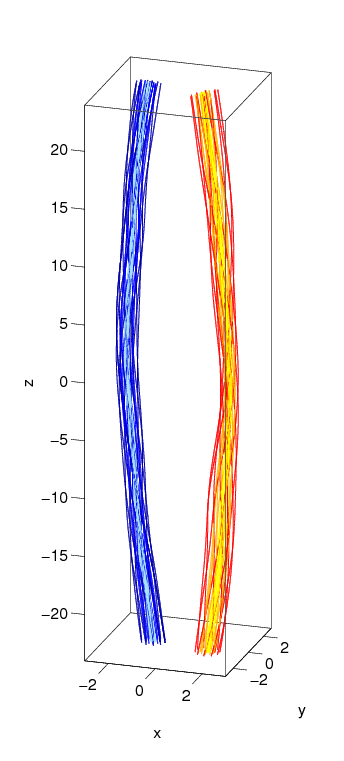}\\
\includegraphics[width=0.24\textwidth]{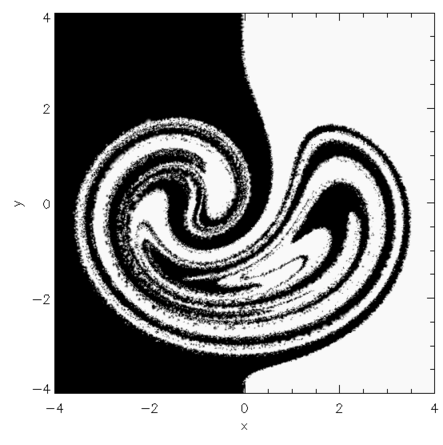}~
\includegraphics[width=0.24\textwidth]{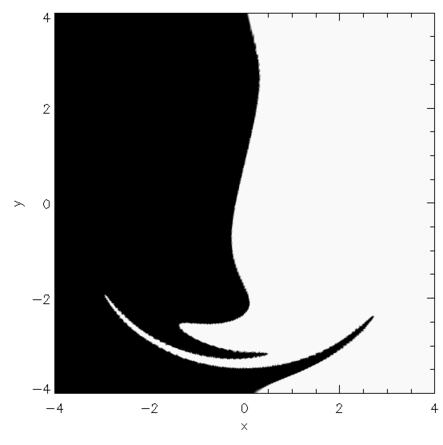}
\end{center}
\caption{Above: field lines traced from fixed locations on the lower boundary (and coloured according 
to location on the lower boundary). Below:  locations of intersection with the plane $z=+24$ of field lines traced from a regular grid in the plane $z=-24$. Field lines traced from locations $x \geq 0$ on the lower boundary are coloured white, from  $x<0$ black. Plots are made for $t=0$ ({\it left}) and $t=290$ ({\it right}) for the run with $\eta=10^{-3}$.}
\label{fig:twofluxtubes}
\end{figure}

The two patches of ${\overline{ \alpha^{*}}}$ shown in Figure~\ref{fig:alpha} are signatures of
the separation of the braided field during the relaxation into a simpler structure consisting of
two weakly twisted magnetic flux tubes of oppositely signed twist, embedded in an approximately uniform field.  Here we provide two simple 
visualizations of the field structure in order to demonstrate this process. 
In Figure~\ref{fig:twofluxtubes} we show sample magnetic field lines in the initial (left) and
final (right) states.  In each case two sets of field lines are shown. First, a set of field lines is traced from three circles of radius 0.1, 0.3 and 0.5 lying in the $z=-24$ plane and centered at 
$\xx_1=(1.415,-0.4875,-24)$. A second set of field lines is traced from three circles of the same radii centered at $\xx_2=(-1.06,0.0375,-24)$.   The centers $\xx_1$ and $\xx_2$ of the two flux tubes have been chosen to lie at the maximum and minimum values of  ${\overline{ \alpha^{*}}}$ (at $z=-24$) in the final state.

In  the lower panels of Figure~\ref{fig:twofluxtubes} a two dimensional representation of this separation is shown.
{We trace field lines from a regular grid of points (over the region $[x,y]\in[\pm6,\pm6]$) on the lower boundary. In the figure we plot the intersections of these field lines with the upper boundary $z=24$, colour-coded such that those field lines that begin at
$x \geq 0$ at $z=-24$ are coloured white, while those with $x<0$ at $z=-24$ are coloured black.  The left panel demonstrates the complexity of the footpoint mapping generated by the braided field in the initial state.}
The right panel shows the much simpler mapping obtained in the final state.

In summary, we have found that in the final state of our simulation
 the magnetic field has two important properties: 
\begin{enumerate}
\item
The magnetic field has been `unbraided', becoming much more topologically simple.
\item
The field relaxes towards a {\it non-linear}, rather than a linear force-free field.
\end{enumerate}

\subsection{Discussion}
The nature of the relaxed state following the {\it resistive} relaxation of astrophysical magnetic fields (such as found in the Solar corona) is an important and unresolved issue. A linear force-free field is often conjectured to be the end-state of  general 
turbulent relaxation processes. This notion was put forward by \cite{taylor1974} to explain the turbulent relaxation of a reversed field pinch (RFP), and was applied to the corona by \cite{heyvaerts1984,heyvaerts1992}.    Noting that the full set of invariants 
of ideal MHD do not hold in a resistive situation, Taylor suggested  that as the magnetic field of an RFP 
decays toward a state of minimum energy, a single invariant -- the total magnetic helicity 
($H = \int \mathbf{A} \cdot \mathbf{B} \ dV$) -- will serve as a constraint on the minimization.  Formulating 
this as a variational problem the end state of relaxation is found to be a linear force-free field,
 which matches experimental data of the RFP rather well \citep[e.g.][]{taylor1986}. In the solar coronal case, on the other hand, it is not yet known whether the total helicity is the only constraint on relaxation. It is certainly true that there are fundamental differences between the the two plasma environments, the most obvious being that field lines in the solar corona do not close there, but rather can be considered to be anchored in the solar photosphere.

In a series of numerical simulations Browning and coworkers  
\citep{browning2003, browning2008, hood2009} have followed the 
development of the kink instability in an initially non-linear force-free cylindrical loop.  In their simulations,  
various initial distributions of $\alpha$ take the form of cylindrically symmetric step-functions with different 
profiles.  The authors find the end-state of relaxation to be a linear force-free field. By contrast,
\cite{amari2000} -- who also tracked a turbulent resistive 
evolution subsequent to a kink instability -- found as an end state a non-linear force-free field containing two distinct flux tubes. 

\cite{hood2009} have suggested that a smoothing of $\alpha$ over the full domain is natural
in the presence of fully developed turbulence, i.e.~with reconnection regions sufficiently spatially 
overlapping (in the case in which a dominant vertical magnetic field $B_z$ is present this implies an overlapping 
of current sheets when projected along field lines onto the $xy$ plane).  However, as we proceed to demonstrate below, our study indeed satisfies such a criterion, though we do not obtain a linear force-free field, as discussed above.  We have recently discovered one additional constraint on the resistive relaxation of line-tied magnetic flux tubes, which demonstrates why a Taylor relaxation to the linear force-free state is prohibited in the braided field studied here \citep{yeates2010}.
It is clear that further studies are necessary to 
determine the end state of relaxation even in isolated flux tubes or braids.

To qualify the above discussion, it is worth emphasising at this point that the final state of our simulations is not an exact force-free field, due to the finite plasma pressure and residual oscillations, as discussed above.
Within each twisted flux tube, there is a balance between magnetic tension associated with their internal twist, and a combination of magnetic and plasma pressure. That the plasma pressure is enhanced in the central part of the domain is not a surprise, since the plasma has been heated there by a multitude of reconnection events. 
Although the Taylor relaxation scenario assumes a very small plasma-$\beta$, and we have used a plasma-$\beta$ of order 0.1, we do not expect that the plasma pressure plays a crucial role in determining the qualitative properties of the final state. We have repeated the simulation run with $\eta=10^{-3}$ with the plasma-$\beta$ reduced by a factor of 10 with no change in the qualitative results.

\section{Magnetic reconnection}\label{recsec}
\subsection{Introduction}
In order for the magnetic field to resistively relax, it is necessary that magnetic reconnection occurs, since a subset of the field lines in the volume have a non-trivial topology -- see Figure \ref{initfig}(a). Reconnection allows a change of the magnetic topology, {and thus release of the energy stored  in the magnetic field}. It is becoming apparent that magnetic reconnection processes are fundamentally different in two dimensions (2D) and three dimensions (3D, clearly the regime applicable to the solar corona). In 2D, the rate at which the magnetic flux is transferred from one flux domain to another (through a magnetic X-point) is given by the value of the electric field evaluated at the null point. However, in 3D, the electric field ${\bf E}$ and magnetic field ${\bf B}$ are no longer perpendicular, and it is the component of ${\bf E}$ parallel to ${\bf B}$ (denoted $E_\|$) that is the crucial quantity. More specifically, magnetic reconnection can be defined to occur within a region in which 
\begin{equation}\label{recrateeq}
F=\int E_\| \, ds \neq 0
\end{equation}
where $s$ is a parameter that runs along magnetic field lines, and where the quantity $E_\|$ should be spatially localised \citep{schindler1988, hesse1988}. The rate of reconnection in this case is given by the maximum value of $F$ over all field lines threading the diffusion region. Thus, the efficiency of the reconnection process is determined not by a local quantity evaluated at a given point, but rather by a `global' property of the field in the vicinity of the diffusion region. Furthermore, the interpretation of this reconnection rate is far more complicated in 3D. The way in which the magnetic flux evolves during the reconnection process has been shown to be fundamentally different in 2D and 3D \citep{priesthornig2003, hornig2003,pontingalsgaard2005}. One crucial difference is that the magnetic field lines change their connectivity continuously throughout the diffusion region. A key result of this property is that magnetic field lines do not reconnect in a one-to-one `cut and paste' fashion, as in simple 2D models. This has a profound effect on the way that the magnetic flux is restructured by the reconnection process. The issue is further complicated by the fact that many different modes of magnetic reconnection may occur in 3D, depending on the local magnetic field structure in the vicinity of the diffusion region, with each having different characteristic behaviour in terms of the evolution of the flux \citep[e.g.][]{hornig2003,pontin2004,pontinhornig2005,wilmotsmith2007a,parnell2010}.

\subsection{Method of measurement}
Thus far we have described the formation of multiple current sheets during our simulations. We now examine the properties of the magnetic reconnection that occurs in these current sheets, in particular the rate of the reconnection. Due to the strong background (`guide') field in the $z$-direction, there are no null points or closed field lines in our volume. The magnetic reconnection takes place in the absence of such topological features, as described for example in the models of \cite{hesse1988,priest1992, hornig2003, wilmotsmith2006, wilmotsmith2009c}. As discussed above, the rate of reconnection is determined, for an isolated diffusion region in 3D,  by the maximum value of $F=\int E_\| \,ds$ attained along any field line threading the diffusion region. 

It is clear from Section \ref{waffle} that there are multiple diffusion regions, at which reconnection is occurring, during the simulations. Therefore the first step towards calculating the reconnection rate is to first identify all of the individual reconnection regions that exist at each time. The global reconnection rate is then obtained by summing the moduli of the maxima of $F$ (see Eq.~\ref{recrateeq}) over all such regions.

\begin{figure}
\centering
\includegraphics[width=0.5\textwidth]{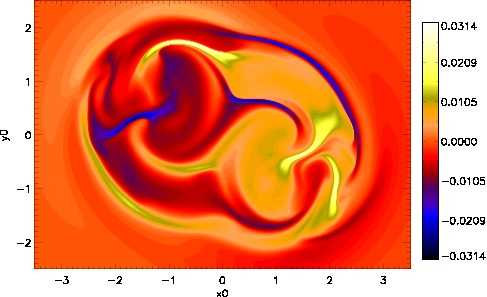}
\caption{Plot of $\Phi(x_0,y_0)$ (as defined in Equation (\ref{phieq})) for $t=50$ for the run with $\eta=10^{-3}$.}
\label{phiall}
\end{figure}
At periodic intervals during the simulations, the data of all physical fields is output, providing a `snapshot' of the evolution. We proceed as follows, treating each individual snapshot in turn. 
\begin{enumerate}
\item
Integrate $E_\|$ along a set of field lines which intersect the $z=0$ plane in a rectangular grid with resolution approximately twice that of the computational mesh (to ensure that no information is lost in areas where the field line mapping shows expansion). For clarity we integrate field lines only over the positive half-space, with the domain having symmetry about the $z=0$ plane. This yields a two-dimensional quantity
\begin{equation}\label{phieq}
\Phi(x_0,y_0)=\int\limits_{s_0}^{s(z=24)} E_\| \, ds 
\end{equation}
subject to $(x,y,z)=(x_0,y_0,0)$ at $s=s_0$, i.e. $(x_0,y_0)$ is the intersection of the appropriate field line with the $z=0$ plane.
\item
Search for the maxima and minima of $\Phi$. We want to count only the reconnection rate for `isolated' diffusion regions, and so we require that $\Phi$ falls below some threshold between peaks in order to class two regions as corresponding to separate reconnection processes. The threshold we choose to class two given maxima as `separate' is that $\Phi$ must drop below 60\% of the lower of the two peak values between the peaks. (In principle we would like $\Phi$ to fall to zero, but  the effect of projecting along the field lines is likely to mean that regions that may be isolated in 3D overlap to some extent in the 2D projection. In addition, numerical quantities are never  exactly zero in practice, so it is not realistic to enforce such a condition.)
\item
To be classed as a reconnection region, maxima/minima are required to contain at least 500 field lines. This corresponds to a cross-sectional area of around 0.02, or around 40 grid cells. (This does not dramatically alter the reconnection rate obtained. However, it discounts large numbers of false positives in the algorithm used to identify isolated maxima, which otherwise selects many peaks which are sufficiently small  (both in spatial coverage and modulus) to be close to the noise threshold.)
\item
To obtain the global reconnection rate, sum the moduli of the identified peak values, and multiply by 2 to obtain a value for the whole domain (since the above procedure is only implemented on the positive half-space in $z$). 
\end{enumerate}
 Below we present the results of implementing this procedure for the standard run with $\eta=10^{-3}$, before going on in the next section to describe the effect of varying the value of $\eta$.
 As a frame of reference for the numerical values obtained for the reconnection rate, we note the following characteristic values in our simulations. Since $\rho\approx 1$ and the equations have been non-dimensionalised by setting $\mu_0 = 1$, we have $|{\bf v}_A| \approx |{\bf B}|\approx 1$. The `guide field' $B_z$ has magnitude of order 1 for all time, and at $t=0$ for example we have peak values of $B_x, B_y$ of approximately 0.3, and rms values of approximately 0.05.
 
 \subsection{Nature and rate of reconnection process}
\begin{figure}
\centering
\includegraphics[width=0.45\textwidth]{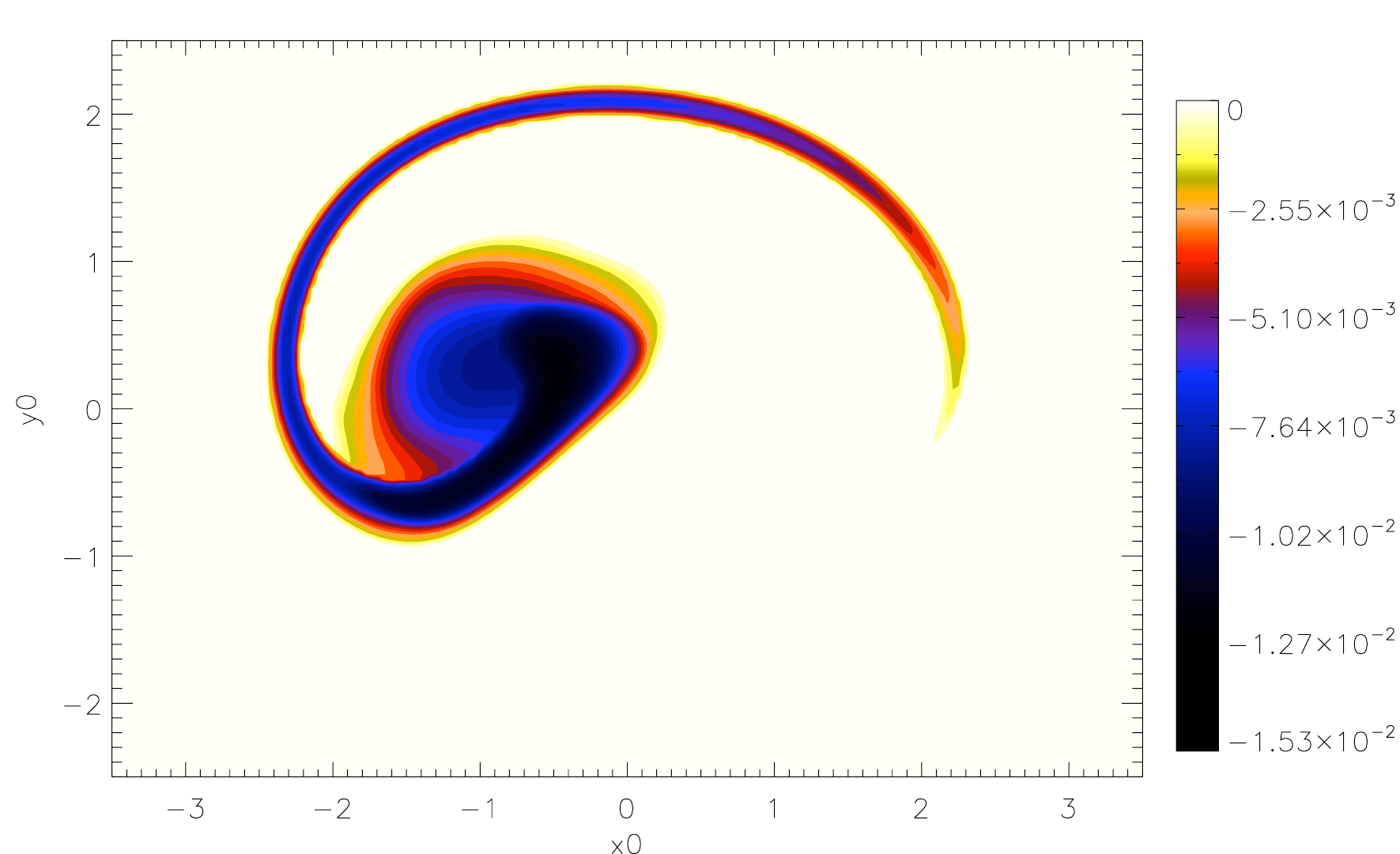}
\includegraphics[width=0.45\textwidth]{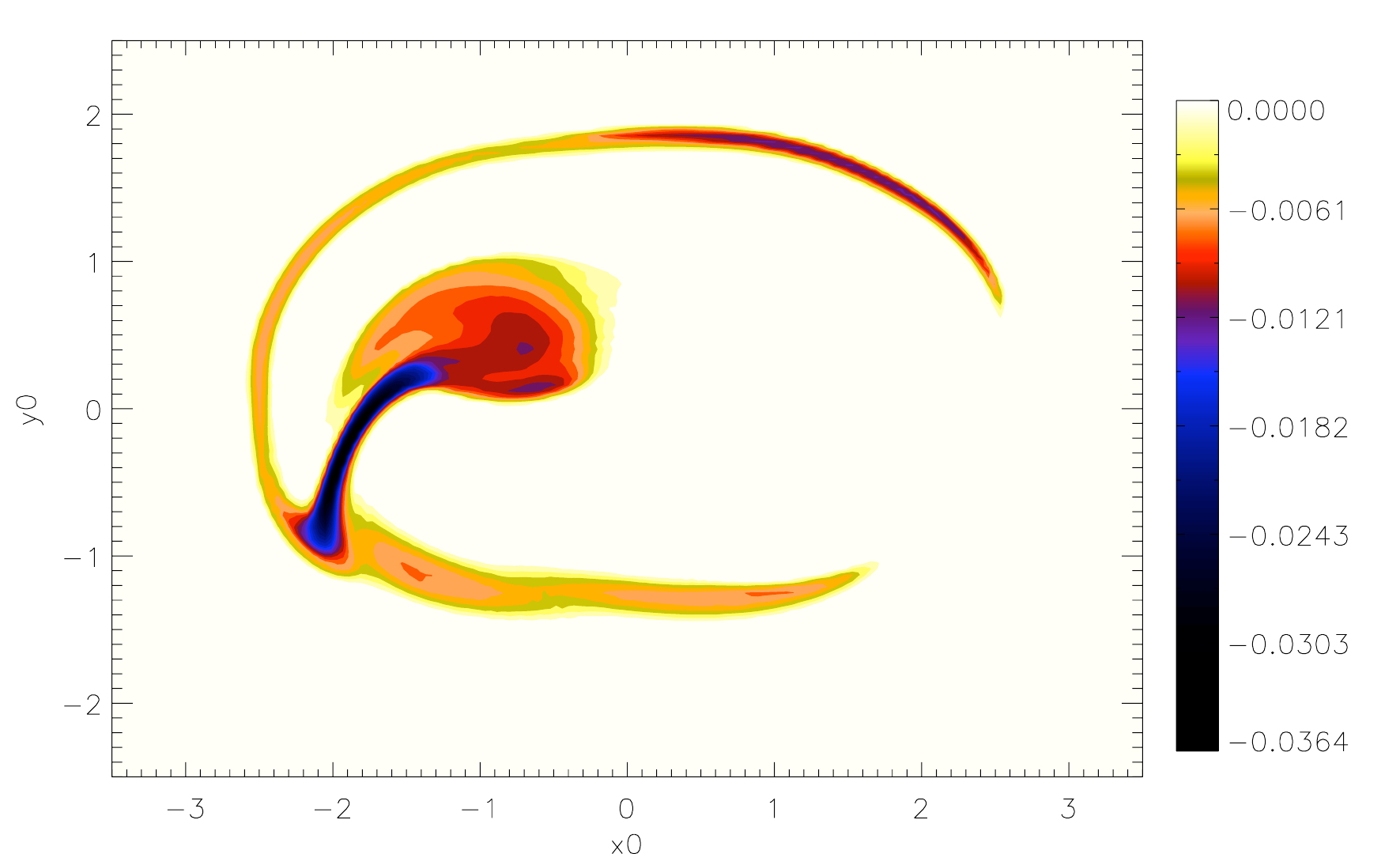}
\includegraphics[width=0.45\textwidth]{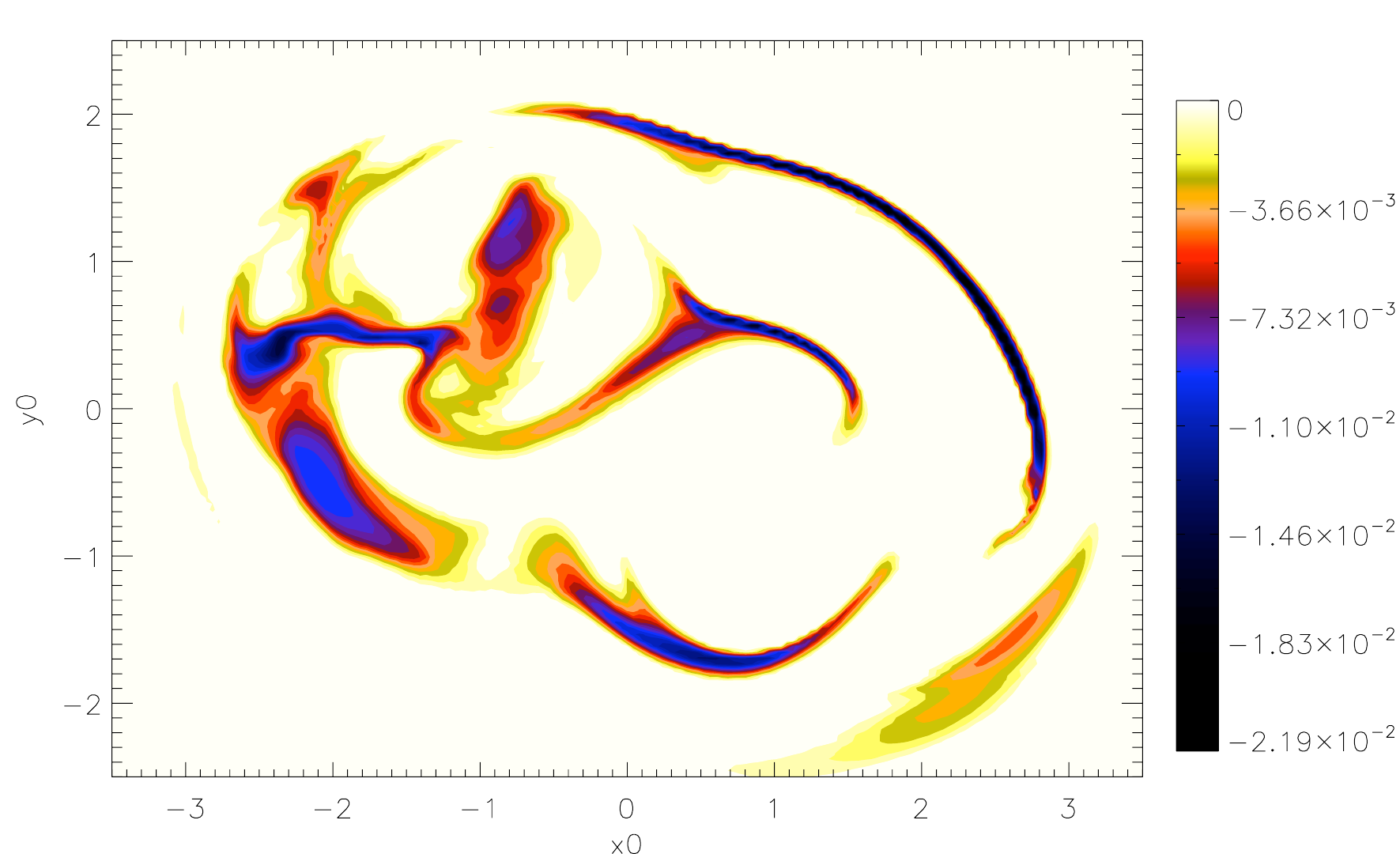}
\includegraphics[width=0.45\textwidth]{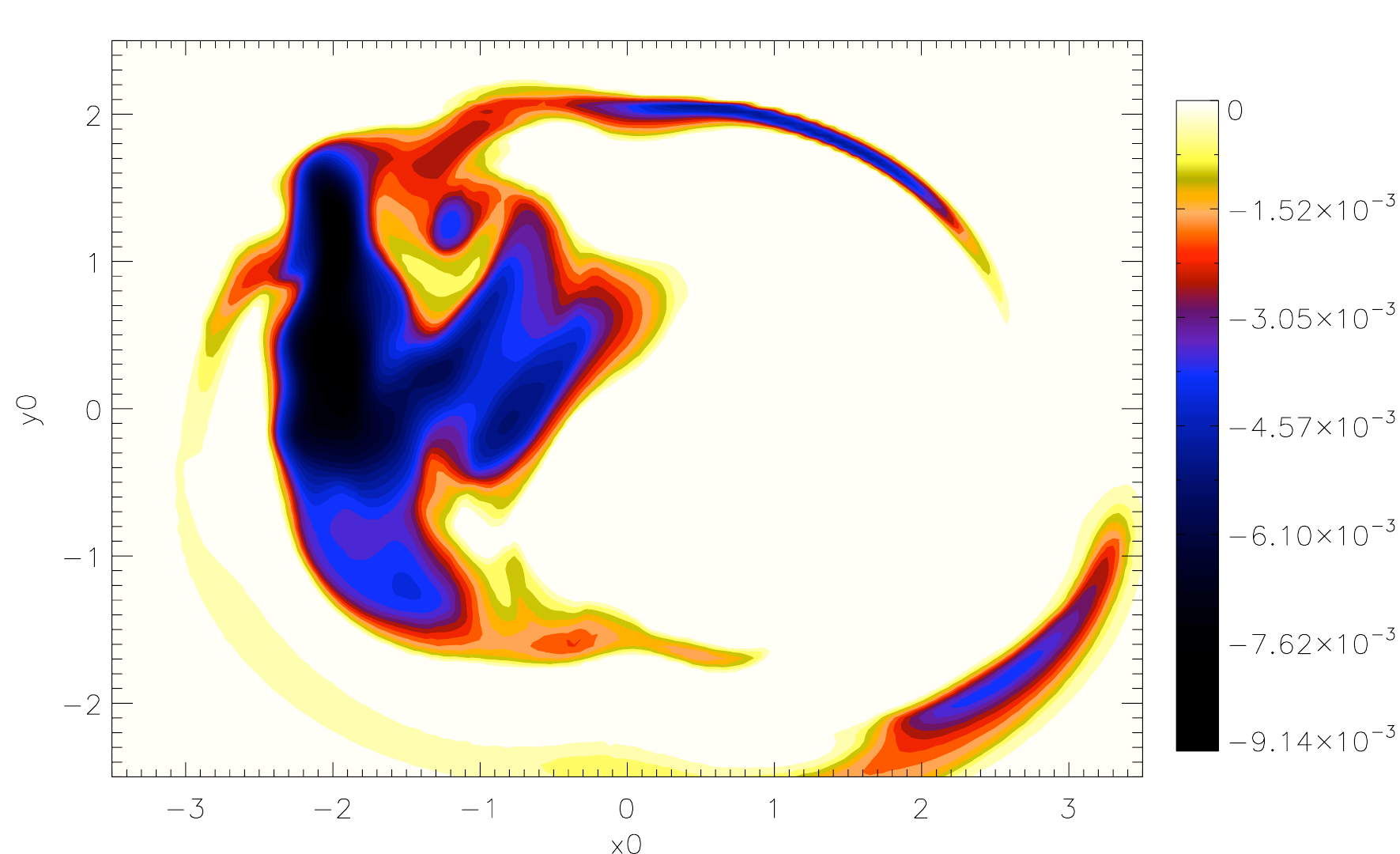}
\includegraphics[width=0.45\textwidth]{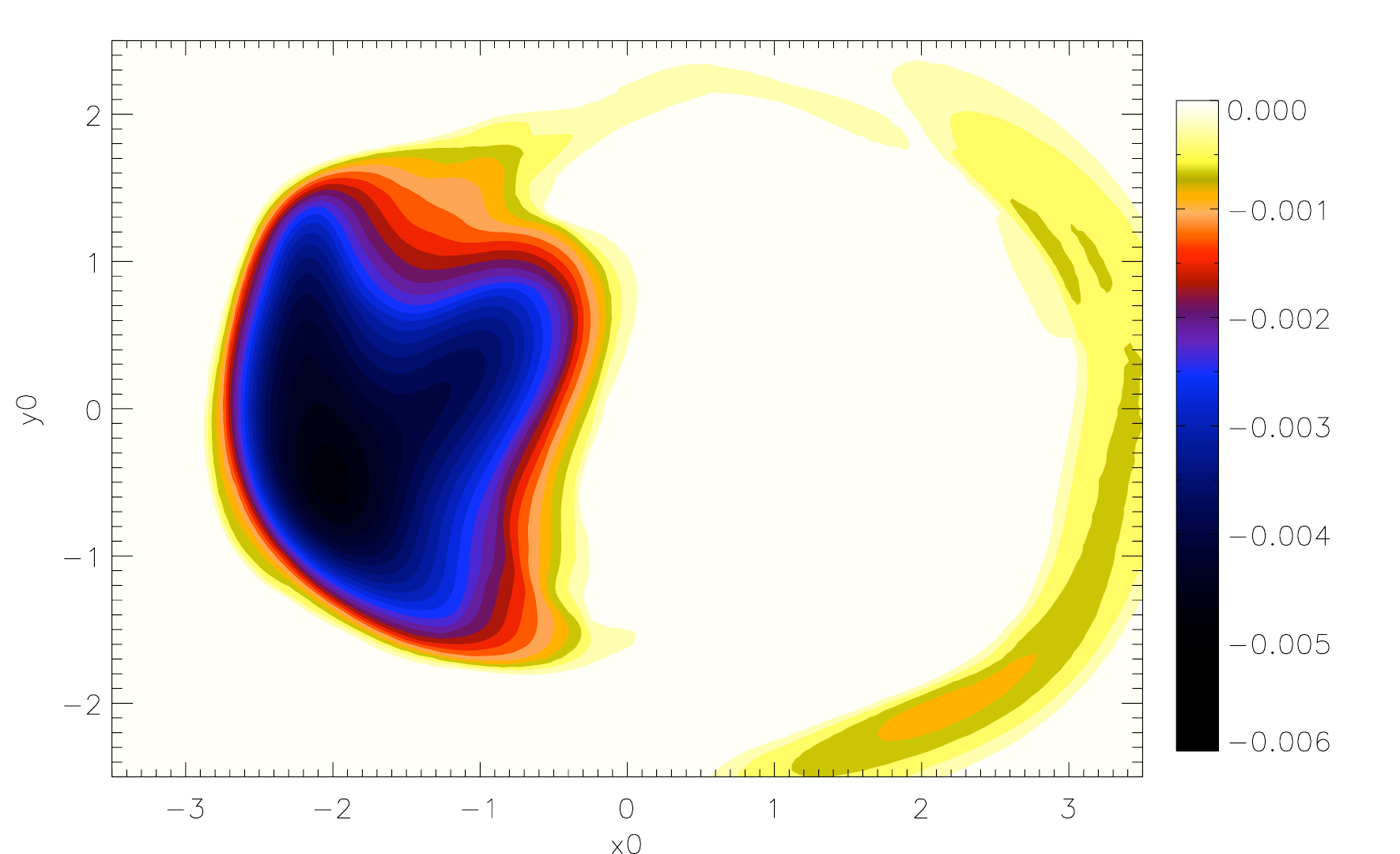}
\caption{Plots of $\Phi$ for $\Phi<0$ only, at times $t=0,15,50,140,290$, for the run with $\eta=10^{-3}$.}
\label{phineg_sd}
\end{figure}

A sample $\Phi$ plot is presented in Figure \ref{phiall}, where the shading intensity shows the value of $\Phi$. It is clear that there exist multiple distinct reconnection regions, which is consistent with the 3D visualisation of the current density presented in Figure \ref{jiso}. We now consider the qualitative evolution of $\Phi$ in time, shown in Figure \ref{phineg_sd}. For clarity only one sign of $\Phi$ is shown ($\Phi<0$). At $t=0$ only a single reconnection region exists, which we see split and then intensify ($t=15$) as time progresses, consistent with the formation of the initial pair of current sheets. As the evolution continues we see the formation of multiple distinct reconnection regions which fill the majority of the central portion of the $x_0 y_0$-plane where the twisted and braided flux is located (this filling is even more apparent when $\Phi$ of both signs is plotted, see Figure \ref{phiall}). As the simulation continues, and the current again assumes a large-scale structure as described above, the number of distinct reconnection regions diminishes once more.

\begin{figure}
\centering
\includegraphics[width=0.49\textwidth]{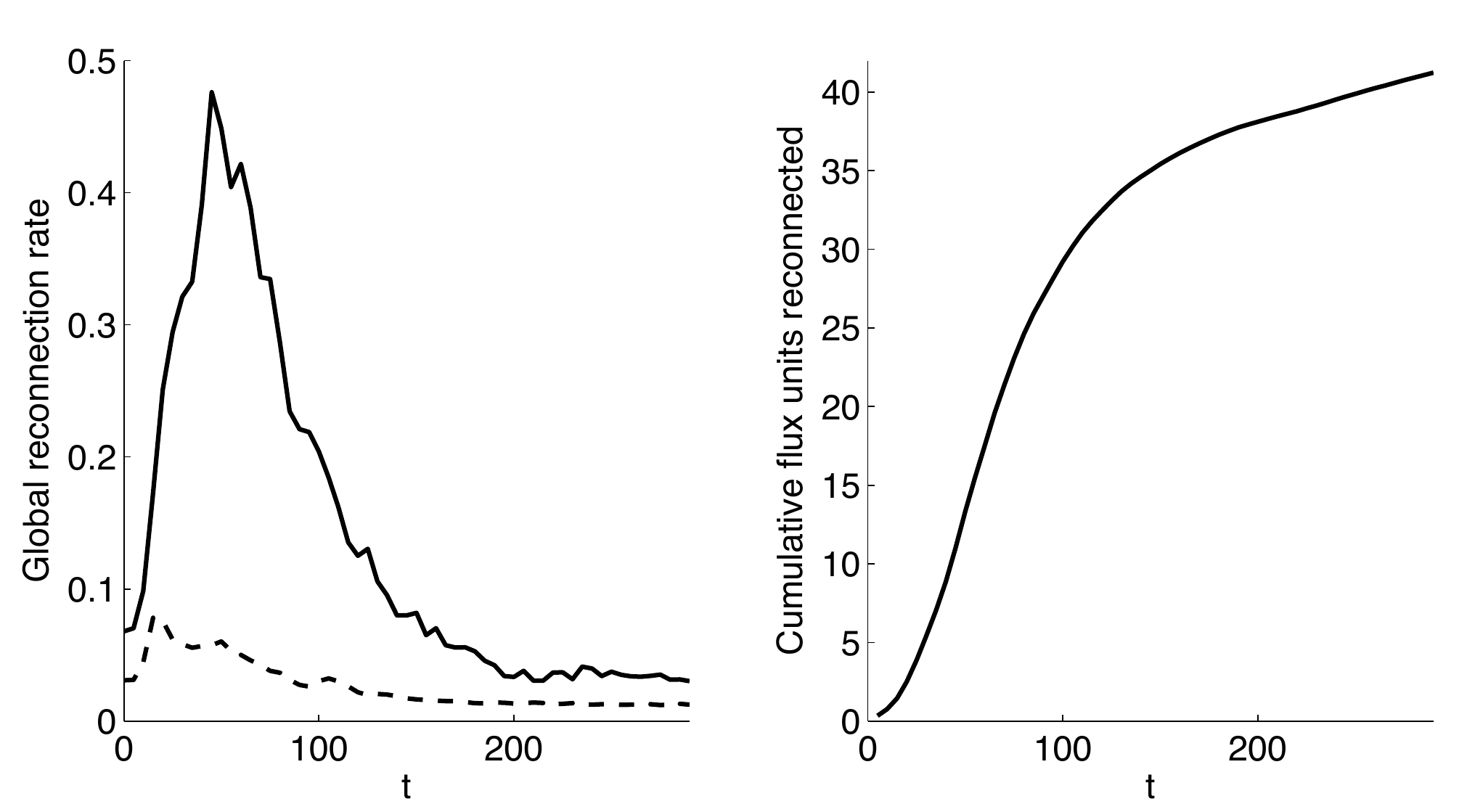}
\caption{Left: Reconnection rate versus time (solid line) for the run with $\eta=10^{-3}$, defined by $2\times\sum|\Phi_{max}|$. Also plotted (dashed line) is $2\times$(absolute maximum of $|\Phi|$). 
Right: Cumulative reconnected flux versus time, for the same run.}
\label{recrate_sd}
\end{figure}

Quantitative measures of the behaviour described above are presented in Figure \ref{recrate_sd}. The global reconnection rate -- plotted as the solid curve in the left panel of Figure \ref{recrate_sd} -- is defined by $2\times\sum|\Phi_{max}|$, as discussed above, where we multiply by 2 to take into account the lower half-space $z<0$.  The plot shows that initially the reconnection rate grows steadily, peaking at $t=35$. There is then a gradual decay until $t=170$ when the rate settles to an approximately constant value (representing large-scale diffusion). The first thing to note is that the time of the peak reconnection rate ($t=35$) does not coincide with the time of the peak current in the domain ($t=15$). Furthermore, it does not occur at the time when the reconnection rate for any given single region is a maximum ($t=15$). Rather, the reconnection rate continues to increase after this time, as a result of the fragmentation of the spatial location of the reconnection process. So the reconnection rate grows {\it not} because of an intensification of the current or an increase of the rate in any {\it one} region, but because of an increase in the fragmentation of the volume in which  reconnection processes take place. This is demonstrated by  the dashed line in the left panel of Figure \ref{recrate_sd}, which shows the maximum rate for the fastest reconnection process (of either sign), defined by $|\Phi|_{global\, max}$, multiplied again by 2 for consistency. The number of individual reconnection regions identified throughout the simulation is shown by the dashed line in the top panel of Figure \ref{rec_compare}.

The reconnection rate as calculated above is of course only an estimation. We would argue that it provides a lower bound on the reconnection rate for the system. There are a number of reasons why it should give a conservative estimate:
\begin{itemize}
\item 
finite spatial resolution of the simulation grid, which will in general underestimate the true current maxima,
\item
finite spatial resolution of the field line grid means the actual peak $\Phi$ will not be exactly obtained,
\item
finite temporal resolution will mean that we fail to capture the highest global peak value,
\item
partial cancellation of positive/negative $E_\|$ along some field lines will degrade some maxima,
\item
exclusion of small regions (step 3 above),
\item
projection along field lines may lead to regions distinct in 3D failing the test for separate regions, criterion described in step 2 above. 
\end{itemize}

\subsection{Reconnected flux}\label{fluxsec}
It is straightforward to estimate the total reconnected magnetic flux from the reconnection rate obtained as described in the previous section, by performing a simple time integral. The cumulative reconnected flux is plotted against time in the right panel of  Figure \ref{recrate_sd}. It turns out that the total number of (non-dimensional) units of flux reconnected during the simulation is $\approx 41.2$. It is revealing to compare this figure with the total poloidal flux in the initial state, which is 30 units. (During the ideal relaxation used to define the initial condition for the present resistive MHD simulations, this quantity is conserved. In the symmetric pre-initial field, which is known in closed form, the poloidal flux is equivalent to the total flux of either sign passing through the $y=0$ plane.) Furthermore, as we saw in Section \ref{finsec}, the final state of the relaxation is not the homogeneous field, but rather contains a finite amount of twist and thus poloidal flux. Due to the high degree of symmetry in the final state, the remaining poloidal flux may be estimated by integrating the positive (say) flux $\BB$ through the $y=0$ plane. The result that we obtain is 15.3 units, indicating that around half of the initial poloidal flux has been cancelled.

To recap then, beginning with 30 units of poloidal flux, 41.2 units are reconnected and 15.3 remain. So the total reconnected flux is greater than the quantity that would appear to be available. The only conclusion is that (at least some of) this flux is reconnected multiple times, most likely in different reconnection regions, during the relaxation and `unbraiding' of the field. On average, each unit of flux is reconnected around 2.8 times. This demonstrates the highly complex nature of the reconnection process that occurs in complex (realistic) magnetic fields such is the one considered here. 

The result should be compared with that of \cite{parnell2008a}, who observed what they termed `recursive reconnection' in a numerical simulation in which two opposite-polarity flux sources were driven past one another in the presence of an overlying field. In their simulation, a number of separator lines (field lines connecting one magnetic null point to another) were formed through which the reconnection took place. Thus the reconnection regions were long-lived and individually identifiable from one time-frame of their simulation to another. They calculated that the magnetic flux was reconnected on average 1.8 times, and argued that this occurred cyclically around the same circuit of reconnection regions (separators). It is clear that we observe a very similar phenomenon occurring here, even though the magnetic field configuration is completely different (and, additionally, isn't continually being driven). The difference in the present case is that due to the lack of topological features present in the magnetic field and the high degree of complexity, it is difficult to label individual reconnection processes and follow them in time. The magnetic flux undergoing  the reconnection process described above then may be best described as being `multiply' rather than `recursively' reconnected.
Given that such similar processes occur in such vastly different configurations, this suggests that such `multiple' or `recursive' reconnection is likely to be ubiquitous in astrophysical magnetic fields.

\section{Dependence on resistivity}\label{etasec}

\begin{figure}
\centering
\includegraphics[width=0.45\textwidth]{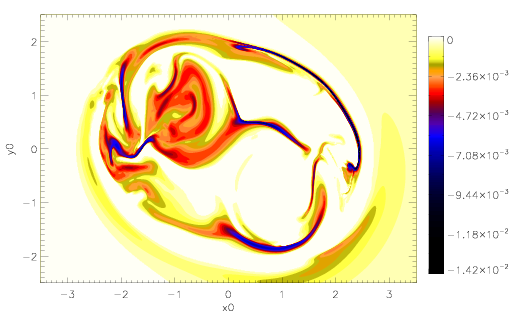}
\includegraphics[width=0.45\textwidth]{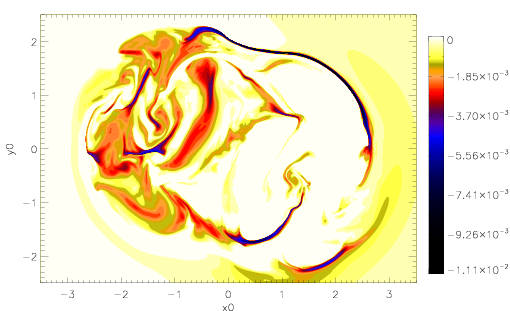}
\includegraphics[width=0.45\textwidth]{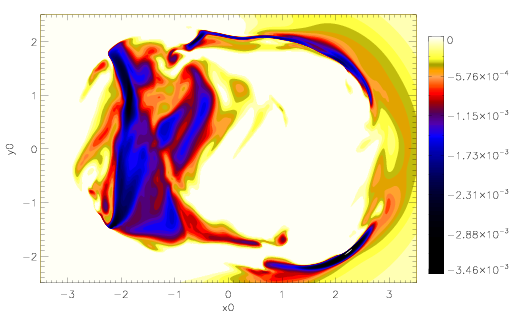}
\caption{Plots of $\Phi$ for $\Phi<0$ only, at times $t=50,140,290$, for the run with $\eta=2\times 10^{-4}$.}
\label{phineg_relax3}
\end{figure}
\begin{figure}
\centering
\includegraphics[width=0.45\textwidth]{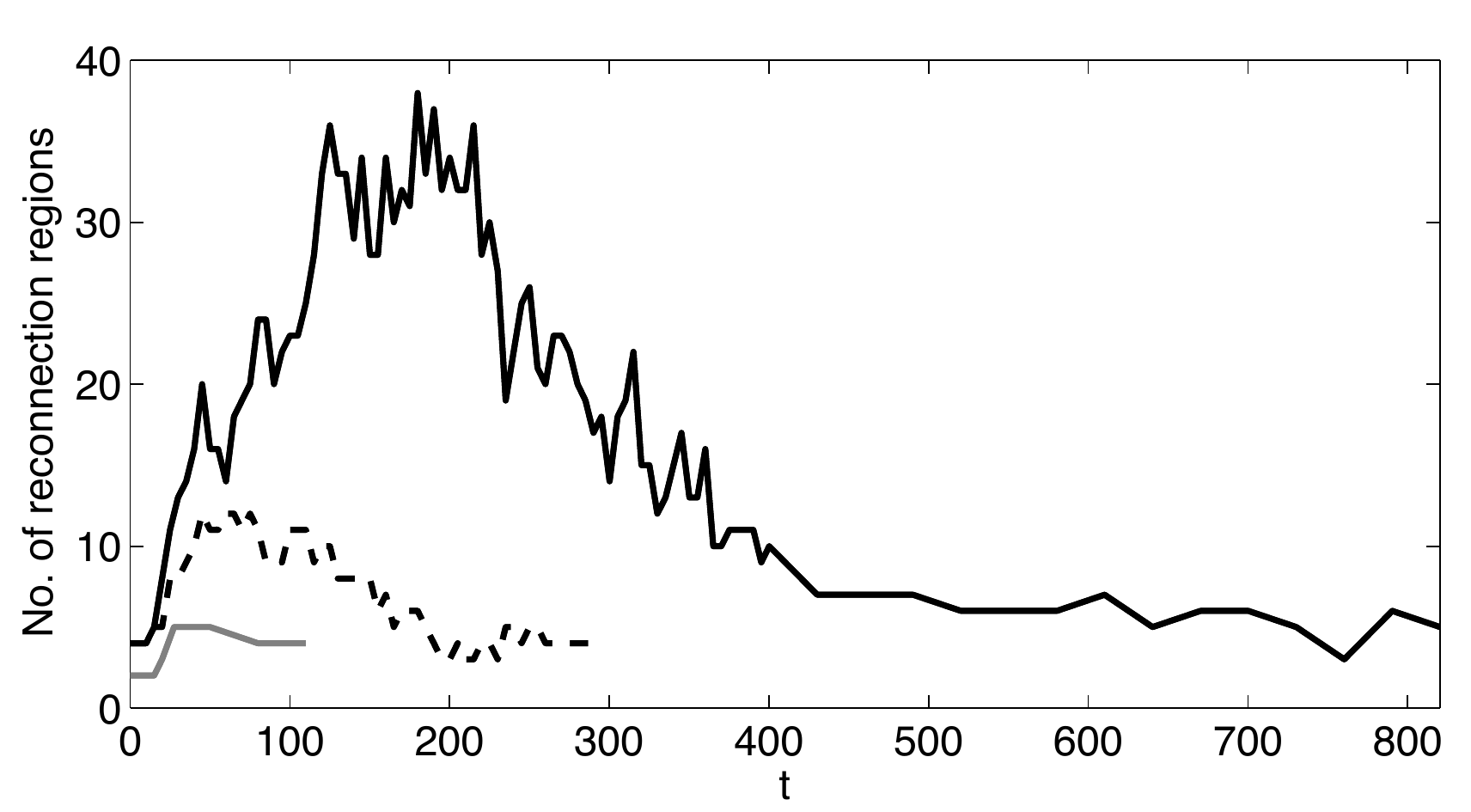}\\ \vspace{0.2cm}
\includegraphics[width=0.45\textwidth]{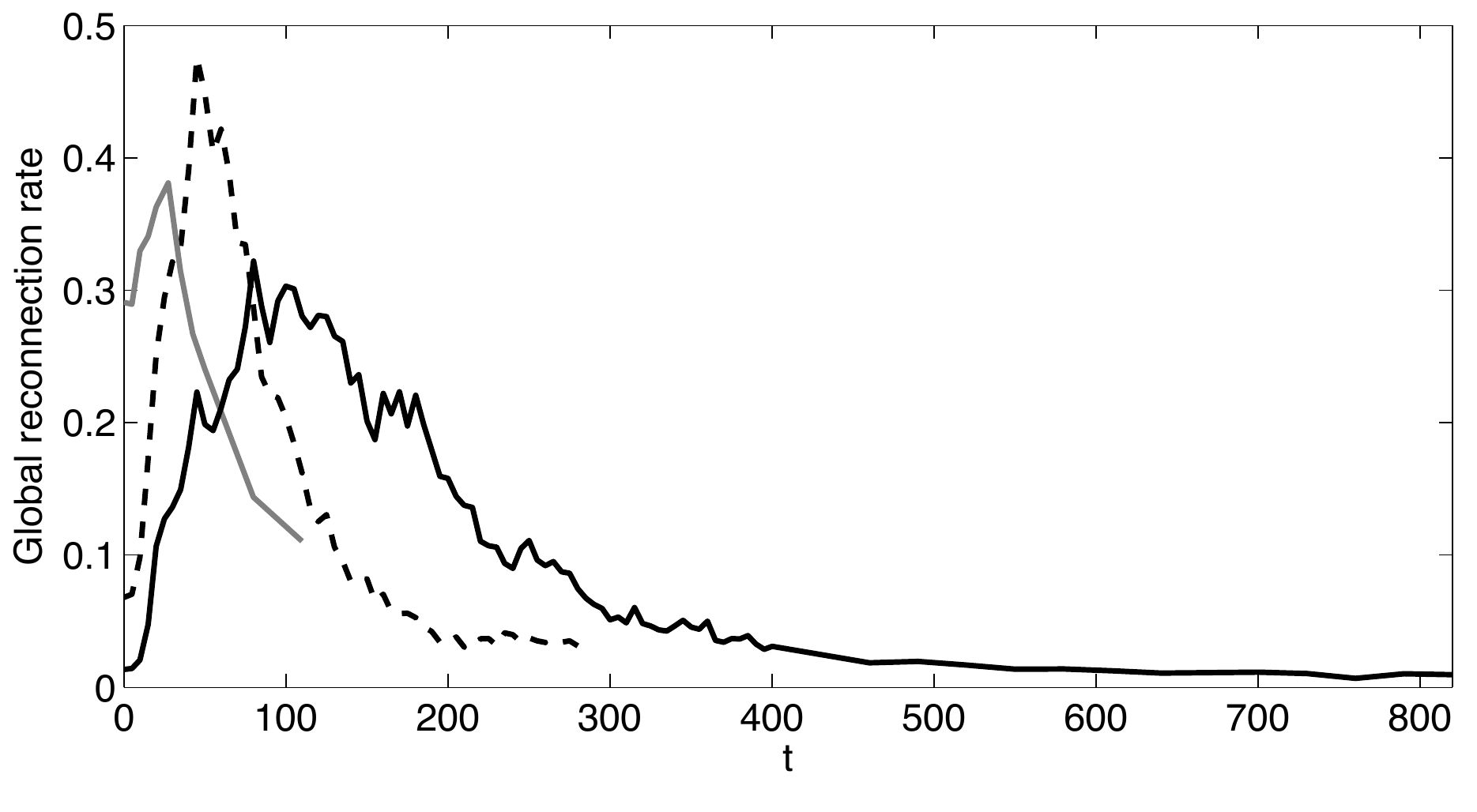}\\ \vspace{0.2cm}
\includegraphics[width=0.45\textwidth]{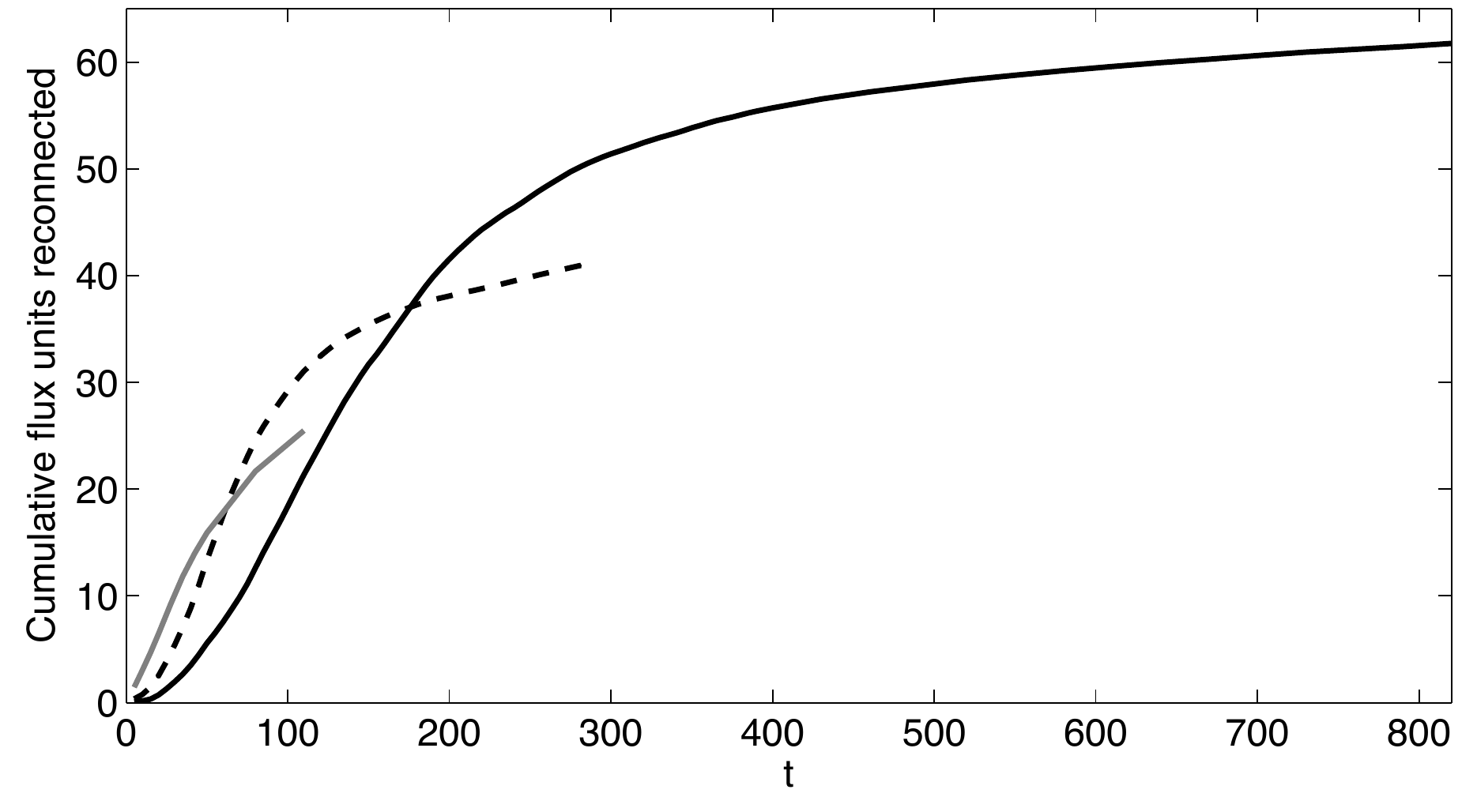}
\caption{Number of identified reconnection regions (upper panel), reconnection rate (middle panel) and cumulative reconnected flux (lower panel) for runs with $\eta=10^{-2}$ (grey solid line), $\eta=10^{-3}$ (black, dashed) and $\eta=2\times 10^{-4}$ (black, solid).}
\label{rec_compare}
\end{figure}
So far we have analysed the resistive relaxation properties of our model braided magnetic field for a fixed value of the resistivity. However, the dissipative length scale in the run we have so far analysed is some orders of magnitude larger than length scales present in the complex field line mapping at $t=0$ \citep[see][]{wilmotsmith2009a}. We therefore proceed in this section to describe the dependence of our relaxation process on the value of the constant resistivity, $\eta$. We have repeated our simulation at one higher and one lower value of $\eta$, namely $\eta=10^{-2}$ and $\eta=2\times 10^{-4}$ (the latter being the lowest value we can use at the present grid resolution while maintaining a reasonable resolution of current sheets in the system).

Qualitatively, the differences between the simulation runs can be described as follows. The initial magnetic field loses stability at a time that is independent of $\eta$, with two current layers forming at locations that are also independent of $\eta$ (see Paper I). However, the subsequent evolution does vary greatly between the simulation runs. First, the lower the value of $\eta$ used, the greater the degree of fragmentation of the current structure. This is quantified below where we also discuss the reconnection rate and evolution of the magnetic flux. It can also be seen by comparison of the frames showing one sign of $\Phi$ in Figures \ref{phineg_sd} and \ref{phineg_relax3}. The second main effect of reducing $\eta$ is to increase the timescale of the relaxation process. {In other words, the plasma resistivity acts to inhibit the tendency to form an increasingly 
complex {system} 
of current layers, such that for larger $\eta$ the system transitions} to a diffusive-timescale evolution in a shorter period of time (again, compare Figures \ref{phineg_sd} and \ref{phineg_relax3}).

The above description can be quantified by analysing the global reconnection rate in the same way as before. First, in Figure \ref{rec_compare} (top panel) we plot the time evolution of the number of individual localised reconnection processes within the domain, for the runs with different $\eta$. It is clear that for lower $\eta$ not only is the peak number higher, but the period over which the reconnection process occurs in a fragmented volume is significantly longer. Of course our algorithm is not sufficiently robust for the exact numbers of regions  identified to have any physical meaning -- however, we are confident that the trends that are demonstrated when varying $\eta$ are physical. Next, in the middle panel of Figure \ref{rec_compare} we compare the measured global reconnection rate versus time. The plots indicate that there is a dependence of reconnection rate on resistivity -- albeit a rather weak one. However, once again the robustness of our algorithm for calculating $\Phi$ could be questioned here. It is also possible that if some small amount of numerical dissipation is present for the low $\eta$ run, then we may under-estimate the reconnection rate by under-estimating $\eta$. 

The greatest insight into the changing nature of the relaxation process is gained by plotting the cumulative reconnected flux versus time for the different simulation runs -- see the lower panel of Figure \ref{rec_compare}. Here we see that the lower $\eta$ becomes, the greater the total amount of flux reconnected in the cascade phase of our simulations. This in itself is not an absolute quantity (there is no precise criterion that we can apply for the transition from a {fast, `reconnective' evolution to a slow} diffusive evolution), but we can compare this number both with the initial poloidal flux and with the net twist remaining in the field at the end of our runs. Note that in the final state of each simulation the field has the same structure as described in Section \ref{finsec} of two oppositely twisted flux tubes. In the run with $\eta=10^{-2}$ our measured cumulative reconnected flux is 24.5 units while 13.1 units remain (compare this with the 30 units of poloidal flux in the initial condition). As discussed in Section \ref{fluxsec}, for $\eta=10^{-3}$ we measure 41.2 units reconnected and 15.3 units remaining. Finally, for the run with $\eta=2\times 10^{-4}$ we measure 61.8 total units reconnected, while 16.2 units remain. We observe that there is a clear trend as $\eta$ is reduced for more reconnection to be required to `unbraid' the field: {the flux is reconnected on average 1.4 times for $\eta=10^{-2}$, 2.8 times for $\eta=10^{-3}$ and 4.4 times for $\eta=2\times 10^{-4}$.} We would expect this trend to continue, at least until $\eta$ has a sufficiently low value that structures on the scale of the initial field line mapping are resolved. As noted by \cite{wilmotsmith2009b}, these length scales decrease exponentially with increasing braid complexity. Therefore in a true braided coronal loop we would expect a highly complex network of current sheets threading the loop, in which the flux is multiply reconnected to a high degree.

\section{Conclusions}\label{concsec}
We have described a series of numerical experiments in which we considered the resistive MHD evolution of a braided magnetic field between two perfectly conducting parallel plates.  Although the magnetic field in the simulations is initially approximately force-free, it was described in Paper I that the field experiences an instability on some characteristic time-scale. Here we considered the subsequent evolution of the system, which is best described as a resistive relaxation. The route taken to find a new equilibrium involves the formation of a complex array of current sheets which are scattered throughout the domain. The current sheets have a ribbon-like appearance, tending to be highly elongated in the direction along the loop, i.e.~parallel to the strong axial field. {The formation of this {myriad} 
of current layers suggests that a turbulent cascade may develop for higher Reynolds numbers than we have been able to use. Due to the absence of boundary forcing, this turbulence would fall under the heading of decaying turbulence.
It would be of interest to push the spatial resolution of our simulations (thus allowing the reduction of $\eta$) to investigate whether a regime of fully-developed turbulence arises, by examining spectral properties of the magnetic and velocity fields. This is beyond the scope of this paper, and we leave it to a future investigation.}

The relaxation process as described above results in a simplification of the magnetic field structure as demonstrated by the mapping of field lines from one line-tied boundary to the other (Figure \ref{fig:twofluxtubes}). The magnetic field lines are untangled, such that no three field lines are braided about one another any longer in the final state. The final state in fact consists of two (weakly) twisted flux tubes, of oppositely signed twist, embedded in an approximately uniform field. This final state approximates a force-free field, $\nabla \times {\bf B}=\alpha\BB$, in which field lines in one twisted flux tube have positive $\alpha$ and in the other negative $\alpha$. Thus the { system approaches} a {non-linear} force-free field, which is not consistent with the Taylor relaxation picture put forward by \cite{heyvaerts1984}. In principle since the net current in the system is zero, a Taylor relaxation would lead to a final state with $\alpha=0$ on every field line, i.e.~the homogeneous potential field. Clearly the single constraint (global helicity conservation) of the Taylor hypothesis is not sufficient to describe this relaxation. Extra constraints on the relaxation which prevent some of the helicity cancellation must be present. 
{Although it is always dangerous to place too much trust in the extrapolation of simulation results to the extremely high Reynolds number coronal plasma,} it appears unlikely that the cause could be related to the `turbulence' {in the system} not being sufficiently developed. In fact we retain {\it more} twist in our flux tubes {in the final state} for lower $\eta$, where the turbulence is better developed. In fact, an additional constraint on the relaxation has recently been discovered \citep{yeates2010}.

The `unbraiding' of the magnetic flux during the relaxation to a non-linear force-free field occurs by magnetic reconnection.  This reconnection occurs in the absence of magnetic nulls or closed field lines (note the presence of a strong background field throughout the domain).  Furthermore, reconnection occurs in a multitude of regions that are spread throughout the volume. In order to determine the efficiency of the reconnection process, we must therefore sum the reconnection rate over all reconnection diffusion regions within the volume. The outcomes of performing such an analysis are as follows:
\begin{enumerate}
\item
The global reconnection rate continues to grow for some time after the peak current in the domain begins to fall. That is, during the intermediate stages of the simulations,  the reconnection rate grows {\it not} because of an intensification of the current or an increase of the rate in any {\it one} region, but rather because of an increase in the fragmentation of the volume in which the reconnection processes take place.
\item
The peak value of the global reconnection rate is at most weakly dependent on the resistivity.
\item
The number of identifiable reconnection regions increases as the resistivity is decreased.
\item
The total quantity of magnetic flux that is reconnected is greater than the total poloidal flux present. This implies that the magnetic flux is `multiply-reconnected' in the complex array of reconnection regions.
\item
As the resistivity is decreased -- {resulting in an increase in the complexity of the 
{multitude}
of current layers} -- the average number of reconnections for each unit of flux increases.
\end{enumerate}

In summary, the loss of stability and subsequent relaxation of braided coronal loops results in a {complex array} of current layers which permit a dissipation of the stored magnetic energy and the attainment of a lower energy non-linear force-free field. 
These results yield a new and intriguing potential resolution to the long-standing debate over the validity of Parker's model of coronal heating by topological dissipation. Specifically, while current singularities are not an inevitable result of random field line tangling, once the field line mapping becomes sufficiently complex, an instability will be triggered, causing the electric current structures to collapse to small scales. The complexity that is inherent in the field then results in {the formation of a fragmented system of current layers} (the sum the associated reconnection processes being `fast' in the sense that the rate of energy release is only weakly dependent on the magnetic Reynolds number). The global result  is a turbulent dissipation of the excess magnetic energy stored in the field, ultimately in the form of heat.
There are many interesting aspects of the above study that warrant further investigation, such as determining what parameters and physical quantities govern the nature of the final state, the possible turbulence properties of the relaxation, and the spatial and temporal distribution of the resulting heating in the loop.

\section*{Acknoweldgements}
{The simulations described were carried out on the DCSC at University of Copenhagen, and on the STFC and SFC (SRIF) funded linux clusters of the UKMHD consortium. 
A.W.S. and G.H. acknowledge financial support from the UK's STFC. }

\bibliographystyle{apalike} 

\end{document}